\documentclass[prd,twocolumn,aps,amsmath,amssymb,nofootinbib,preprintnumbers]{revtex4}
\usepackage{graphicx}
\usepackage{bm}
\usepackage{amsmath}
\usepackage{amsfonts}
\usepackage{braket}
\usepackage{amssymb}
\usepackage{epstopdf}
\usepackage{hyperref}
\usepackage{mathrsfs}
\usepackage{subfigure}
\usepackage{ulem}

\begin{document}
\newcommand{\drawsquare}[2]{\hbox{%
\rule{#2pt}{#1pt}\hskip-#2pt
\rule{#1pt}{#2pt}\hskip-#1pt
\rule[#1pt]{#1pt}{#2pt}}\rule[#1pt]{#2pt}{#2pt}\hskip-#2pt
\rule{#2pt}{#1pt}}

\newcommand{\Yfund}{\raisebox{-.5pt}{\drawsquare{6.5}{0.4}}}
\newcommand{\Yasymm}{\raisebox{-3.5pt}{\drawsquare{6.5}{0.4}}\hskip-6.9pt%
                     \raisebox{3pt}{\drawsquare{6.5}{0.4}}%
                    }
\newcommand{\Ysymm}{\Yfund\hskip-0.4pt%
                    \Yfund}
\def\symm{\Ysymm}
\def\bsymm{\overline{\Ysymm}}
\def\ls{\mathrel{\lower4pt\vbox{\lineskip=0pt\baselineskip=0pt
           \hbox{$<$}\hbox{$\sim$}}}}
\def\gs{\mathrel{\lower4pt\vbox{\lineskip=0pt\baselineskip=0pt
           \hbox{$>$}\hbox{$\sim$}}}}
\def\drawbox#1#2{\hrule height#2pt
        \hbox{\vrule width#2pt height#1pt \kern#1pt
              \vrule width#2pt}
              \hrule height#2pt}

\def\Fund#1#2{\vcenter{\vbox{\drawbox{#1}{#2}}}}
\def\Asym#1#2{\vcenter{\vbox{\drawbox{#1}{#2}
              \kern-#2pt       
              \drawbox{#1}{#2}}}}
\def\sym#1#2{\vcenter{\hbox{ \drawbox{#1}{#2} \drawbox{#1}{#2}    }}}
\def\fund{\Fund{6.4}{0.3}}
\def\asymm{\Asym{6.4}{0.3}}
\def\bfund{\overline{\fund}}
\def\basymm{\overline{\asymm}}


\newcommand{\beq}{\begin{equation}}
\newcommand{\eeq}{\end{equation}}
\def\ls{\mathrel{\lower4pt\vbox{\lineskip=0pt\baselineskip=0pt
           \hbox{$<$}\hbox{$\sim$}}}}
\def\gs{\mathrel{\lower4pt\vbox{\lineskip=0pt\baselineskip=0pt
\def\lsim{\mathrel{\lower4pt\vbox{\lineskip=0pt\baselineskip=0pt
           \hbox{$<$}\hbox{$\sim$}}}}
\def\gsim{\mathrel{\lower4pt\vbox{\lineskip=0pt\baselineskip=0pt
           \hbox{$>$}\hbox{$\sim$}}}}           \hbox{$>$}\hbox{$\sim$}}}}


\title{Probing  the Supersymmetric Inflaton and Dark Matter link via the CMB, LHC and  XENON1T experiments}

\author{C\'eline B\oe hm $^{1,2}$}
\author{Jonathan Da Silva $^{1,2}$}
\author{Anupam Mazumdar~$^{3,4}$}
\author{Ernestas Pukartas~$^{3}$}

\affiliation{$^{1}$~Institute for Particle Physics Phenomenology, University of Durham, Durham, DH1 3LE, UK}
\affiliation{$^{2}$~LAPTh, U. de Savoie, CNRS,  BP 110,
  74941 Annecy-Le-Vieux, France}
\affiliation{$^{3}$~Consortium for Fundamental Physics, Lancaster University, LA1 4YB, UK}
\affiliation{$^{4}$~Niels Bohr Institute, Copenhagen University, DK-2100, Denmark}

\begin{abstract}
The primordial inflation dilutes all matter except the quantum fluctuations which we see in the cosmic microwave background (CMB) radiation. Therefore the last phases of inflation must be embedded within a beyond the Standard Model (SM) sector where the inflaton can directly excite the SM quarks and leptons. In this paper we consider two inflaton candidates $\widetilde{L}\widetilde{L}\widetilde{e}$ and $\widetilde{u}\widetilde{d}\widetilde{d}$ whose decay can naturally excite all the relevant degrees of freedom besides thermalizing the lightest supersymmetric particle (LSP) during and after reheating. In particular, we present the regions of the parameter space which can yield successful inflation with the right temperature anisotropy in the CMB, the observed relic density for the neutralino LSP, and the recent Higgs mass constraints from LHC within the MSSM with non-universal Higgs masses -- referred to as the NUHM2 model. We found that in most scenarios, the LSP seems strongly mass degenerated with 
the next 
to lightest LSP (NLSP) and the branching ratio $B_s \rightarrow \mu^+ \mu^-$  very close to the present bound, thus leading to falsifiable predictions. Also the dark matter interactions with XENON nuclei would fall within the projected range for the XENON1T experiment. In the case of a positive signal of low scale supersymmetry at the LHC, one would be able to potentially pin down the inflaton mass by using the associated values for the mass of the stau, the stop and the neutralino.
\end{abstract}

\maketitle

\section{Introduction}

The primordial inflation must explain the seed perturbations for the cosmic microwave background (CMB) radiation~\cite{WMAP},
and after the end of inflation the coherent oscillations of the inflation must excite the Standard Model (SM) quarks and leptons at temperatures sufficiently high to realize baryons and dark matter in the current universe~\cite{Mazumdar:2010sa,Mazumdar:2011zd}. In this respect, it is vital that the last phase of primordial inflation must end in a vacuum of Beyond the Standard Model (BSM) physics which can solely excite the relevant degrees of freedom required for the success of Big Bang Nucleosynthesis (BBN), see for a review~\cite{BBN}.

Inflation needs a potential which remains sufficiently flat along which the slow-roll inflation can take place in order to generate the observed temperature anisotropy in the CMB. The low scale supersymmetry (SUSY) guarantees the flatness of such flat directions at a perturbative and a non-perturbative level (for a review see~\cite{Enqvist:2003gh}), besides providing a falsifiable framework for the BSM physics, see~\cite{BSM}. Furthermore, the lightest SUSY particle can be absolutely stable under R-parity, and thus provides an ideal cold dark matter candidate~\cite{CDM}.

The flat directions of SUSY, especially the MSSM provides nearly 300 gauge-invariant $F$-and $D$-flat directions 
~\cite{Gherghetta:1995dv,Dine:1995kz}, which are all charged under the SM gauge group. Out of these flat directions, there are particularly 2 $D$-flat directions: $\widetilde{u}\widetilde{d}\widetilde{d}$ and $\widetilde{L}\widetilde{L}\widetilde{e}$, which carry the SM charges and  can be ideal inflaton candidates~\cite{Allahverdi:2006iq,Allahverdi:2006we,Allahverdi:2006cx}. Here $\widetilde u,~\widetilde d$ correspond to the right handed squarks, $\widetilde L$ corresponds to the left handed slepton, and $\widetilde e$ corresponds to the right handed (charged) leptons.  Both the inflaton candidates provide an {\it inflection point} in their respective potentials where inflation can be driven for sufficiently large e-foldings of inflation to explain the current universe and explain the seed perturbations for the temperature anisotropy in the CMB~\cite{Allahverdi:2006iq,Allahverdi:2006cx}. 

The inflaton in this case only decays into the MSSM degrees of freedom which thermalize the universe with a temperature, $T_R\sim 10^{8}$~GeV~\cite{Allahverdi:2011aj}. This temperature is sufficient to excite the degrees of freedom which are needed for the lightest supersymmetric particle (LSP) to get a relic density that matches observations. It is then natural to ask whether there exists any parameter space, where both successful inflation and thermal dark matter abundance can be explained simultaneously~\cite{Allahverdi:2007vy,Allahverdi:2010zp}~\footnote{Inflationary models embedded within a hidden sector with SM gauge singlets suffer a serious drawback -- it is not at all clear why and how such an inflaton would decay {\it solely} into the SM degrees of freedom. A hidden sector inflaton can couple to many other 
hidden sectors of the BSM, given the fact that any stringy construction of BSM produces large number of hidden sectors within landscape, which can in principle accommodate the Kaluza-Klein dark matter as a candidate~\cite{Douglas:2004zg}. The top-down construction of inflation generically excites the hidden sectors predominantly as compared to the visible sector fields, see~\cite{Cicoli}.}.

Recently both the ATLAS and CMS experiments have 
recorded hints of a Higgs boson. With an integrated luminosity of respectively $4.8$ and $4.9$ fb$^{-1}$ and a  centre of mass energy of $\sqrt{s}=7$~TeV the ATLAS experiment reported 
an excess of events in the $H\rightarrow ZZ^*\rightarrow 4l$ (where $l$ is either electrons or muons) and the $H\rightarrow\gamma\gamma$ channels \cite{atlas1,atlas2}. In addition, ATLAS published a broad excess in the $H\rightarrow WW^* \rightarrow l \nu l' \nu$ channel from a combined analysis with these two channels \cite{ATLAS:2012ae}. All these signals would point out towards a $\sim$ 125 GeV Higgs boson while also excluding masses outside the [116,131] GeV range (a part from a small window between 238 and 251 GeV). In addition to these results, CMS also observed an excess of events but pointing towards a Higgs mass of 
$\sim 119$~GeV, which was not observed by the ATLAS collaboration.

In this paper we explore scenarios which lead to a Higgs mass in the  allowed range, a correct relic density for the neutralino LSP and inflaton properties which are in agreement with the CMB data. We use a variant of the MSSM, which differs from the constrained MSSM and minimal supergravity model (mSUGRA)~\cite{mSUGRA}.
Recently, it was shown indeed that the specific value of $m_h\simeq 125$ GeV is hard to accommodate within mSUGRA which has a rather restrictive parameter space with degenerate scalar masses $m_0$, and gaugino masses $m_{1/2}$, at the grand unification (GUT) scale~\cite{Baer,Baer:2011ab,Ellis:2012aa}. For high values of $m_0$ and $m_{1/2}$ it is possible to get mSUGRA regions with right dark matter abundance~\cite{Akula:2011aa,Feng:2012jf}, but they may not be accessible at the LHC scale. It was pointed out nevertheless that a variant of MSSM with non-universal Higgs masses~\cite{ref19 of Ellis paper} known as NUHM2 can accommodate both the $119$ and $125$ GeV Higgs mass values and the observed relic density for the LSP. It is then natural to ask whether, in this framework, the same set of parameters leads to an inflaton mass which is compatible with the CMB observations and the current particle physics spectrum at low energies.

Since in our case the inflaton candidates are gauge invariant, by using the renormalization group equations (RGE) at one loop level, one can evaluate the mass of the inflaton, $m_\phi$, from the scale of inflation to the scale of LHC. This eventually will enable us to relate the inflaton mass with the  dark matter parameter space and the CP-even Higgs mass.  

The plan of this paper is to briefly discuss in Section II the properties of inflation and various observables of CMB, which can be satisfied by the two flat direction candidates, $\widetilde u\widetilde d\widetilde d$ and $\widetilde L \widetilde L \widetilde e$. We will also discuss how these two candidates can also generate departure from random gaussian fluctuations which can be verified by the forthcoming satellite experiment Planck~\cite{Mazumdar:2011xe}. In section III, we identify  benchmark points where the constraints on inflation and the LSP relic density converge towards a Higgs mass of about $119$ GeV or $125$ GeV. 
In section IV, we perform a broader scan of the NUHM2 parameters in order to delineate the regions of the parameter space where the neutralino has the observed relic density and the Higgs mass falls within the allowed range. In the final section we will discuss how LHC observables and also direct dark matter searches (in particular the XENON1T~\cite{Aprile:2005mz} experiment) can be used to probe the NUHM2 parameter space and eventually pin down the inflaton mass.


\section{Inflation, CMB observables \& Renormalization Group Equations}
\label{ICM}
 
\subsection{Inflaton Candidates: Flat Directions of Squarks and Sleptons}
\label{MSSMI}


\begin{figure}[t]
\centering
\includegraphics[width=0.90\linewidth]{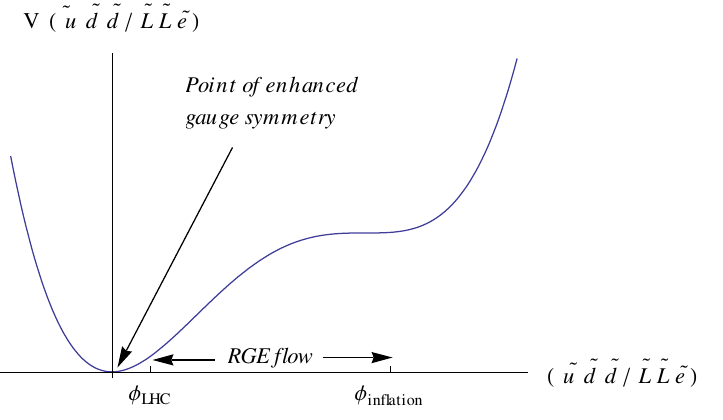}
\caption{A schematic drawing of inflationary potential for either $\widetilde u\widetilde d\widetilde d$ or $\widetilde L\widetilde L\widetilde e$ as shown in Eq.~(\ref{scpot}). Inflation happens near the {\it inflection} point as shown by $\phi_{\rm inflation}=\phi_0$, inflation ends at the point of {\it enhanced gauge symmetry}, where the entire (MS)SM gauge symmetry is recovered. The physical mass and couplings at high scale $\phi_{0}$ and $\phi_{LHC}$ are related via RGE described by Eqs.~(\ref{rgudd},\ref{rglle}).
}\label{potential}
\end{figure}


In Refs.~\cite{Allahverdi:2006iq,Allahverdi:2006we,Allahverdi:2007vy,Allahverdi:2010zp} the authors have recognized two $D$-flat directions which can be the ideal inflaton candidates, because both $\widetilde{u}\widetilde{d}\widetilde{d}$ and $\widetilde{L}\widetilde{L}\widetilde{e}$ flat directions are lifted by higher order superpotential terms of the following form which would provide non-vanishing $A$-term in the potential even at large VEVs~\footnote{~Note that the R-parity is still conserved, both the superpotentials: ${\bf udd}$ and ${\bf LLe}$ do not appear at the renormalizable level, they are instead lifted by non-renormalizable operators, see Refs.~\cite{Allahverdi:2006iq,Allahverdi:2006we,Allahverdi:2007vy,Allahverdi:2010zp} for a detailed discussion. Further note that both the operators vanish in the vacuum which is shown as $\phi=0$ in Fig.~\ref{potential}. Rest of the MSSM flat directions are lifted by hybrid operators of type $W \supset \Phi^{n-1}\Psi/M_{\rm P}^{n-3}$ where $\Phi$ and $\Psi$ 
are MSSM superfields, see~\cite{Dine:1995kz,Gherghetta:1995dv, Enqvist:2003gh}. Such operators do not induce non-renormalizable $A$-term during inflation which is relevant for inflation, see Refs.~\cite{Allahverdi:2006iq,Allahverdi:2006we} for discussion.}:
\beq \label{supot}
W \supset {\lambda \over 6}{\Phi^6 \over M^3_{\rm P}}\, ,
\eeq
 where $\lambda \sim {\cal O}(1)$~\footnote{~The exact value of $\lambda$ is irrelevant for the CMB analysis, as it does not modify the CMB predictions. However it is possible to extract its value by integrating out the heavy degrees of freedom. If the origin of these operators arise from either $SU(5)$ or $SO(10)$, then the typical value is of order $\lambda \sim {\cal O}(10^{-2})$ for $SO(10)$ and $\lambda \sim {\cal O}(1)$ for $SU(5)$, as shown in Ref.~\cite{Allahverdi:2007vy}.}. The scalar component of $\Phi$ superfield, denoted by $\phi$, is given by~\footnote{The representations for the flat directions are given by:
$\widetilde u^{\alpha}_i=\frac1{\sqrt{3}}\phi\,,~
\widetilde d^{\beta}_j=\frac1{\sqrt{3}}\phi\,,~
\widetilde d^{\gamma}_k=\frac{1}{\sqrt{3}}\phi .$
Here $1 \leq \alpha,\beta,\gamma \leq 3$ are color indices, and $1
\leq i,j,k \leq 3$ denote the quark families. The flatness constraints
require that $\alpha \neq \beta \neq \gamma$ and $j \neq k$.
$\widetilde L^a_i=\frac1{\sqrt{3}}\left(\begin{array}{l}0\\ \phi\end{array}\right)\,,~
\widetilde L^b_j=\frac1{\sqrt{3}}\left(\begin{array}{l}\phi\\ 0\end{array}\right)\,,~
\widetilde e_k=\frac{1}{\sqrt{3}}\phi\,,$
where $1 \leq a,b \leq 2$ are the weak isospin indices and $1 \leq
i,j,k \leq 3$ denote the lepton families. The flatness constraints
require that $a \neq b$ and $i \neq j \neq k$. Note that the cosmological perturbations do not care which combination arises, as gravity couples universally.
}
\beq \label{infl}
\phi = {\widetilde{u} + \widetilde{d} + \widetilde{d} \over \sqrt{3}} ~ ~ ~ , ~ ~ ~ \phi = {\widetilde{L} + \widetilde{L} + 
\widetilde{e} \over \sqrt{3}},
\eeq
for the $\widetilde{u}\widetilde{d}\widetilde{d}$ and $\widetilde{L}\widetilde{L}\widetilde{e}$ flat directions respectively.
After minimizing the potential along the angular direction $\theta$ ($\Phi$ = $\phi e^{i \theta}$), we can situate the real part of $\phi$ by rotating it to the corresponding angles $\theta_{\rm min}$. The scalar potential is then found to be~\cite{Allahverdi:2006iq,Allahverdi:2006we}
\beq \label{scpot}
V(\phi) = {1\over2} m^2_\phi\, \phi^2 - A {\lambda\phi^6 \over 6\,M^{3}_{\rm P}} + \lambda^2
{{\phi}^{10} \over M^{6}_{\rm P}}\,,
\eeq
where $m_\phi$ and $A$ are the soft breaking mass and the $A$-term respectively ($A$ is a positive quantity since its phase is absorbed by a redefinition of $\theta$ during the process)~\footnote{Note that the supergravity corrections do not spoil the inflationary potential. Typically supergravity corrections lead to Hubble induced mass corrections, but in our case the Hubble parameter during inflation is always much smaller than $m_{\phi}$, see Eq.~(\ref{hubble}), for a detailed discussion see Refs.~\cite{Allahverdi:2006iq,Allahverdi:2006we}.}.
The masses for $\widetilde{L}\widetilde{L}\widetilde{e}$ and $\widetilde{u}\widetilde{d}\widetilde{d}$ are given by:
\begin{eqnarray}\label{masses}
m^2_{\phi}=\frac{m^2_{\widetilde L}+m^2_{\widetilde L}+m^2_{\widetilde e}}{3}\,,\\
m^2_{\phi}=\frac{m^2_{\widetilde u}+m^2_{\widetilde d}+m^2_{\widetilde d}}{3}\,.
\end{eqnarray}
These masses are now VEV dependent, i.e. $m^2(\phi)$. The inflationary perturbations will be able to constrain 
the inflaton mass only at the scale of inflation, i.e. $\phi_0$, while LHC will be able to constrain the masses at the LHC scale. However both the physical 
quantities are related to each other via RGE as we will discuss below.
For
\beq \label{dev}
{A^2 \over 40 m^2_{\phi}} \equiv 1 - 4 \alpha^2\, ,
\eeq
where $\alpha^2 \ll 1$, there exists a point of inflection ($\phi_0$) in $V(\phi)$~\footnote{The value of $\alpha$ during inflation could be 
small, i.e. $\alpha \sim 10^{-10}$, but it runs dynamically from the GUT scale where $A^2=40m_{\phi}^2$ to the required value at scale of inflation via the RG-equations. For a detailed discussion see Ref.~\cite{Allahverdi:2010zp}.}
, where
\begin{eqnarray}
&&\phi_0^4 = {m_\phi M^{3}_{\rm P}\over \lambda \sqrt{10}} + {\cal O}(\alpha^2) \, , \label{infvev} \\
&&\, \nonumber \\
&&V^{\prime \prime}(\phi_0) = 0 \, , \label{2nd}
\end{eqnarray}
at which
\begin{eqnarray}
\label{pot}
&&V(\phi_0) = \frac{4}{15}m_{\phi}^2\phi_0^2 + {\cal O}(\alpha^2) \, , \\
\label{1st}
&&V'(\phi_0) = 4 \alpha^2 m^2_{\phi} \phi_0 \, + {\cal O}(\alpha^4) \, , \\
\label{3rd}
&&V^{\prime \prime \prime}(\phi_0) = 32\frac{m_{\phi}^2}{\phi_0} + {\cal O}(\alpha^2) \, .
\end{eqnarray}
From now on we only keep the leading order terms in all expressions. 
Note that inflation occurs within an interval~\footnote{For a low scale inflation, setting the initial condition is always challenging. However in the case of a MSSM or string theory landscape where there are many false vacua at high and high scales, then it is conceivable that earlier phases of inflation could have occurred in those false vacua. This large vacuum energy could lift the flat direction condensate
either via quantum fluctuations~\cite{Allahverdi:2007wh}, however see also the challenges posed by the quantum fluctuations~\cite{Enqvist:2011pt}, or via classical initial condition which happens at the level of background without any problem, see~\cite{Allahverdi:2008bt}.}
\beq \label{plateau}
\vert \phi - \phi_0 \vert \sim {\phi^3_0 \over 60 M^2_{\rm P}} ,
\eeq
in the vicinity of the point of inflection, within which the slow roll parameters $\epsilon \equiv (M^2_{\rm P}/2)(V^{\prime}/V)^2$ and $\eta \equiv M^2_{\rm P}(V^{\prime \prime}/V)$  are smaller than $1$. The Hubble expansion rate during inflation is given by
\beq \label{hubble}
H_{inf} \simeq \frac{1}{\sqrt{45}}\frac{m_{\phi}\phi_0}{M_{\rm P}}\,.
\eeq
In order to obtain the flat potential, it is crucial that the $A(\phi_0)$-term ought to be close to $m_{\phi}(\phi_0)$ in the above potential Eq.~(\ref{scpot}). This can be obtained within two particular scenarios:

\begin{itemize}

\item {\it Gravity Mediation}: \\
In gravity-mediated SUSY breaking, the $A$-term and the soft SUSY breaking mass are of the same order of magnitude as the gravitino mass, i.e. $m_{\phi} \sim A $~\cite{Nilles:1983ge}. 

\item{\it Split SUSY}: \\ 
Normally in Split SUSY scenario where the scale of SUSY is high and sfermions are very heavy, the $A$-term is typically protected by R-symmetry, see Refs.~\cite{ArkaniHamed:2004fb,Giudice:2004tc}, as a result the $A$-term could be very small compared to the soft masses. However, if the Yukawa hierarchy arises from the Froggatt-Nielsen mechanism, then the $A$-term can be as large as that of the soft mass, i.e. $m_\phi \sim A$, as in the case of Ref.~\cite{Babu:2005ui}.

\end{itemize}

Keeping low scale and high scale SUSY breaking scenarios in mind here we will consider a large range of $(m_{\phi},~\phi_0)$ to match the cosmological observations.


\subsection{Cosmological Observables}


\subsubsection{Gaussian fluctuations and tensor to scalar ratio}
The above potential Eq.~(\ref{scpot}) has been studied extensively in Refs.~\cite{Allahverdi:2006we,Bueno Sanchez:2006xk,Enqvist:2010vd}. The amplitude of density perturbations $\delta_H$ and the scalar spectral index $n_s$ are given by:
\beq \label{ampl}
\delta_H = {8 \over \sqrt{5} \pi} {m_{\phi} M_{\rm P} \over \phi^2_0}{1 \over \Delta^2}
~ {\rm sin}^2 [{\cal N}_{\rm COBE}\sqrt{\Delta^2}]\,, \eeq
and
\beq \label{tilt}
n_s = 1 - 4 \sqrt{\Delta^2} ~ {\rm cot} [{\cal N}_{\rm COBE}\sqrt{\Delta^2}], \eeq
respectively, where
\beq \label{Delta}
\Delta^2 \equiv 900 \alpha^2 {\cal
N}^{-2}_{\rm COBE} \Big({M_{\rm P} \over \phi_0}\Big)^4\,. \eeq
In the above, ${\cal N}_{\rm COBE}$ is the number of e-foldings between the time when the observationally relevant perturbations are generated till the end of inflation and follows: 
${\cal N}_{\rm COBE} \simeq 66.9 + (1/4) {\rm ln}({V(\phi_0)/ M^4_{\rm P}}) \sim 50$. 
Since the perturbations are due to a single field, one does not expect large non-Gaussianity from this model ($f_{NL}\leq 1$, see Ref.~\cite{Maldacena}). 

\begin{figure}[t]
\includegraphics[width=1.0\linewidth]{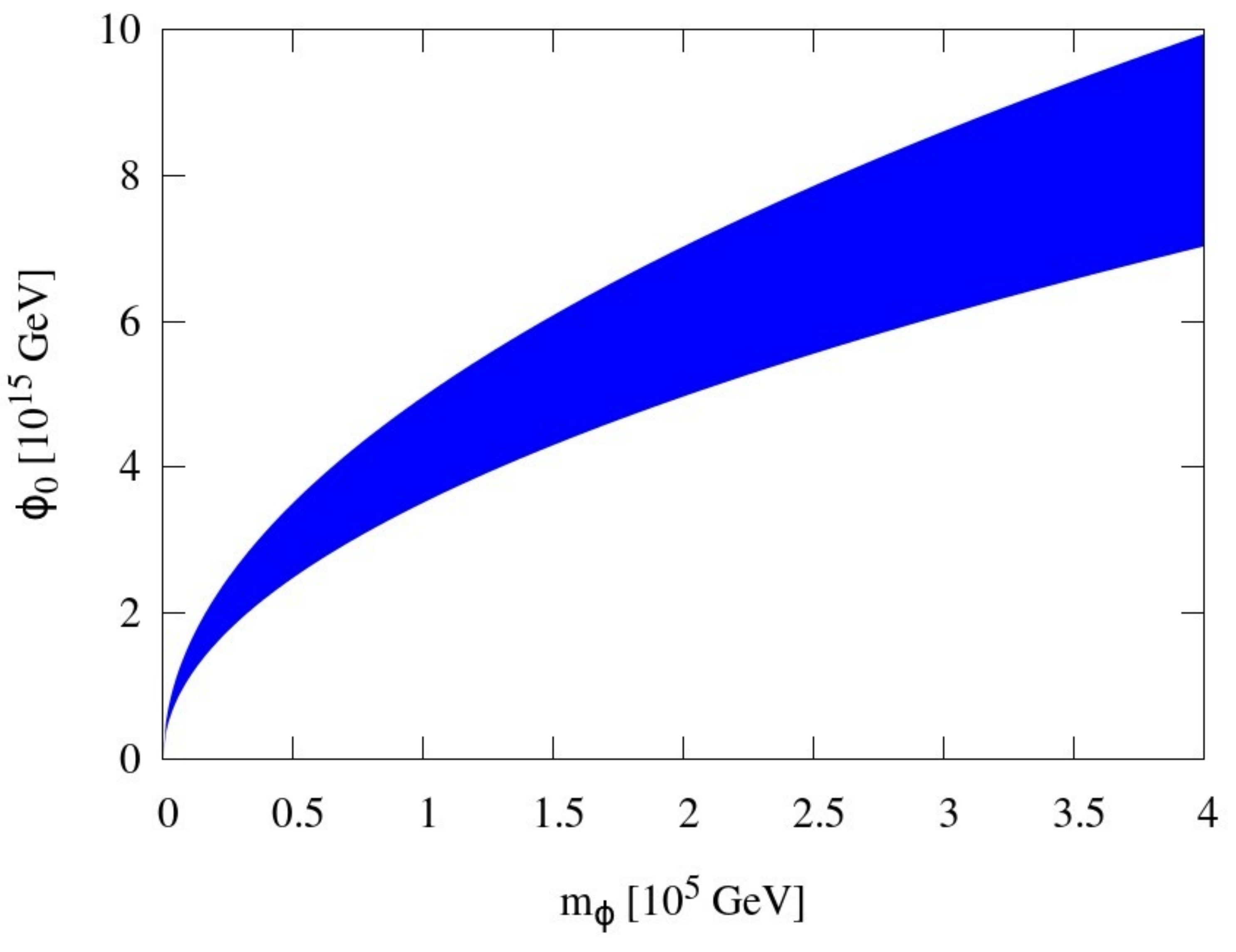}
\caption{$(\phi_0,m_{\phi})$ plane in which inflation is in agreement with the cosmological observations of the temperature anisotropy of the CMB fluctuations. The blue region shows the inflaton energy scale and inflaton mass which are compatible with the central value of the amplitude of the seed perturbations, $\delta_H=1.91\times 10^{-5}$, and the $2\sigma$ allowed range of spectral tilt 
$0.934 \leq n_s\leq 0.988$~\cite{WMAP}. 
Note that we restricted ourselves to inflaton VEVs $\phi_0$ below 
the GUT scale.}\label{Fig-0}
\end{figure}

In Fig.~\ref{Fig-0} we have explored a wide range of the inflaton mass, $m_{\phi}$, where inflation can explain the observed 
temperature anisotropy in the CMB with the right amplitude,
 $\delta_H=1.91\times 10^{-5}$, and the tilt in the power spectrum, $0.934 \leq n_s\leq 0.988$~\cite{WMAP}. Fig.\ref{Fig-0} represents the inflation energy scale versus the mass of the inflaton. The configurations which fit the observed values of  
 $\delta_H$ and $n_s$ are shown in blue. Although we have restricted ourselves to VEV values below the GUT scale, the model does provide  negligible running in the tilt which is well within the observed limit.

Here we have allowed for a wide range of $m_{\phi}$ and $\phi_0$ values because ultimately we want to show that inflation can happen within low-scale SUSY scenarios from high-scale SUSY breaking soft-masses (cf the split-SUSY scenario
~\cite{Babu:2005ui}). 

In this paper we will mostly consider scenarios where the scale of inflation is low enough that one would not expect any observed tensor perturbations in any future CMB experiments. To obtain large observable tensor to scalar ratio, $r$, one would have to embed these inflaton candidates within $N=1$ supergravity (SUGRA). This would modify the potential with a large vacuum energy density besides providing 
SUGRA corrections to mass and A-term~\cite{Mazumdar:2011ih,Hotchkiss:2011gz}. One could then obtain  
$r\sim 0.05$ for both inflaton flat directions: $\widetilde u\widetilde d\widetilde d$ and $\widetilde L\widetilde L\widetilde e$~
as shown in \cite{Hotchkiss:2011gz}.

\subsubsection{Non-Gaussianity: interplay between $\widetilde u\widetilde d\widetilde d$ and $\widetilde L\widetilde L\widetilde e$}

Note that within MSSM flat directions, $\widetilde u\widetilde d\widetilde d$ and $\widetilde L\widetilde L\widetilde e$ are two independent directions which are lifted by themselves. In principle both the flat directions can be lifted by higher order terms. One can imagine that either of $\widetilde u \widetilde d\widetilde d$ or $\widetilde L\widetilde L\widetilde e$ are lifted by higher order superpotential operators in Eq.~(\ref{supot}), such as  $\Phi^9,~\Phi^{12}, \cdots$, see~\cite{Allahverdi:2011su}, while the other is lifted at $\Phi^6$. This will create a hierarchy in potential energies between the two flat directions. One will have a large vacuum energy density compared to the other.

This second flat direction which is lifted by lower order operators is sometimes known as a curvaton in the literature, see for a review \cite{Mazumdar:2010sa}.
Therefore, either $\widetilde u\widetilde d\widetilde d$ or $\widetilde L\widetilde L\widetilde e$ could be the inflaton or the curvaton depending on term in the superpotential which lifts them. In this case the curvaton typically slow rolls and decays later, which is responsible for generating a sizable non-Gaussianity in the squeezed-limit, which is given by~\cite{Mazumdar:2011xe} and can be detectable by the Planck satellite:
\begin{equation}
f_{NL}^{squezed}\sim {\cal O}(1)h^{2/3}\sim {\cal O}(5)-{\cal O}(1000)\,.
\end{equation}
for $h\sim 10^{-1}-10^{-5}$, where $h$ denotes the SM Yukawa couplings. The  $f_{NL}$ depends on a particular decay channel of
the squarks and sleptons, therefore the Yukawa couplings appear in the analysis.  In fact the smallest Yukawa coupling dominates $f_{NL}$.
In this scenario the perturbations are mainly created by the flat direction which is lifted at the lowest order in the superpotential.
When there are two flat directions which are lifted simultaneously, the  thermalization process is delayed a lot due to a phenomenon known as {\it kinematical blocking} of the decay channels~\cite{Allahverdi:2005mz}. This explains the non-Gaussianity. However, as we shall see 
the rest of our analysis  will be independent of whether significant non-Gaussianity is generated or not after inflation.


\subsubsection{Reheating, thermalization, \& thermal history}

Instant reheating and thermalization occurs when a single flat direction is responsible for inflation and structure formation. 
This is due to the gauge couplings of the inflaton to gauge/gaugino fields. Within $10-20$ inflaton oscillations radiation-dominated universe prevails, as shown in Ref.~\cite{Allahverdi:2011aj}. The resultant reheat temperature at which all the MSSM {\it degrees of freedom} are in thermal equilibrium (kinetic and chemical equilibrium) is given by~\cite{Allahverdi:2011aj}
\begin{equation}
T_{rh}\sim 2\times 10^{8}~{\rm GeV}.
\end{equation}
Since the temperature of the universe is so high, it immediately thermalizes the LSP provided it has gauge interactions. The LSP relic density is then given by the Standard (thermal) Freeze-out mechanism.
In particular, if the neutralino is the LSP, its relic density is determined by its annihilation and coannihilation rates. 

The advantage of realizing inflation in the visible sector is that it is possible to {\it nail} down the thermal history of the universe precisely. 
At temperatures below $10-100$~GeV there will be no extra degrees of freedom in the thermal bath except that of the SM, therefore BBN can proceed without any trouble within low scale SUSY scenario. This reheat temperature is marginally compatible with the BBN bound for the gravitino mass $m_{3/2}\geq {\cal O}(\rm TeV)$. It is also sufficiently high that various mechanisms of baryogenesis may be invoked to generate the observed baryon asymmetry of the universe.


\subsection{Renormalization Group Equations}

Since the inflaton carries the SM charges and they are fully embedded within the MSSM, it is possible to probe various regions of the parameter space for inflation. The CMB fluctuations probe the inflaton potential at the inflationary scale. At low energies the inflaton
properties can be probed by the LHC from the masses of the squarks and sleptons.

The inflaton mass and the non-renormalizable $A$ term in the inflationary potential are both scale dependent quantities, and they can be tracked down to lower energies by using the RGE. In \cite{Allahverdi:2006we,Allahverdi:2007vy,Allahverdi:2010zp}, it was shown that at one loop level for the relevant flat direction, $\widetilde u\widetilde d\widetilde d$:
\begin{equation}
\begin{aligned}
\label{rgudd}
&\hat{\mu} \frac{dm^2_\phi}{d\hat{\mu}}=-\frac{1}{6\pi^2}(4M_3^2g_3^2+\frac{2}{5}M_1^2g_1^2),
\\&\hat{\mu} \frac{dA}{d\hat{\mu}}=-\frac{1}{4\pi^2}(\frac{16}{3}M_3g_3^2+\frac{8}{5}M_1g_1^2).
\end{aligned}
\end{equation}
where  $\hat{\mu} = \hat\mu_0=\phi_0$ is the VEV at which inflation 
occurs. For $\widetilde{L}\widetilde{L}\widetilde{e}$:
\begin{equation}
\begin{aligned}
\label{rglle}
&\hat\mu\frac{dm^2_\phi}{d\hat\mu}=-\frac{1}{6\pi^2}(\frac{3}{2}M_2^2g_2^2+\frac{9}{10}M_1^2g_1^2),
\\&\hat\mu\frac{dA}{d\hat\mu}=-\frac{1}{4\pi^2}(\frac{3}{2}M_2g_2^2+\frac{9}{5}M_1g_1^2),
\end{aligned}
\end{equation}
where $M_1$, $M_2$, $M_3$ are $U(1)$, $SU(2)$ and $SU(3)$ gaugino masses, which all equate to $m_{1/2}$ at the unification scale, and $g_1$, $g_2$ and $g_3$ are the associated couplings. To solve these equations, one needs to take into account of the running of the gaugino masses and coupling constants which are given by, see \cite{Nilles:1983ge}:
\begin{equation}
\beta (g_i)=\alpha_i g_i^3 \hspace{1.5cm}
\beta\bigg{(}\frac{M_i}{g_i^2}\bigg{)}=0,
\end{equation}
with $\alpha_1={11}/{16\pi^2}$, $\alpha_2={1}/{16\pi^2}$ and $\alpha_1=-{3}/{16\pi^2}$. So every point in $(m_0,m_{1/2})$ 
(where $m_0$ and $m_{1/2}$ denote the scalar masses and the gauginos at the unification scale  respectively)
plane can now be mapped onto $(\phi_0,m_\phi)$ plane~\footnote{The RGE equations also exhibit explicitly that the fine-tuning required to match $m_\phi$ and the $A$-term at the {\it inflection point} can be obtained from the running of the gauge couplings, see Ref.~\cite{Allahverdi:2010zp}. At the LHC scale the ratio of soft masses and the $A$-term is order one.}.


\section{NUHM2 scenario \& constraining the NUHM2 parameter space}

The NUHM2 is a variant of MSSM with non-universal soft breaking masses $m_1$ and $m_2$ which are independent for both Higgs doublets ~\cite{Ellis1, Ellis:2002iu, Ellis:2002wv}. The universality of scalar masses $m_0$ at the unification scale, i.e. GUT scale, is still assumed, but in NUHM2 model, they are different from $m_1$ and $m_2$. It is well-known that the Higgs masses can be written as, see~\cite{Ellis:2002wv, Ellis:2002iu}:
\begin{equation}
\begin{aligned}
&m_1^2(1+\tan^2\beta)=M_A^2\tan^2\beta-\mu^2(\tan^2\beta+1-\Delta_\mu^{(2)})\\&-(c+2c_\mu)\tan^2\beta-\Delta_A\tan^2\beta-\frac{1}{2}m_Z^2(1-\tan^2\beta)-
\Delta_\mu^{(1)}
\end{aligned}
\end{equation}
and
\begin{equation}
\begin{aligned}
m_2^2(1+\tan^2&\beta)=M_A^2-\mu^2(\tan^2\beta+1+\Delta_\mu^{(2)})\\&-(c+2c_\mu)-\Delta_A+\frac{1}{2}m_Z^2(1-\tan^2\beta)+\Delta_\mu^{(1)},
\end{aligned}
\end{equation}
where $c$, $c_\mu$, $\Delta_\mu^{(1,2)}$, $\Delta_A$ are radiative corrections, $\mu$ -- Higgs mixing parameter, $M_A$ is the mass of CP-odd pseudo-scalar Higgs and 
$m_Z$ is the mass of the $Z$ boson.

In fact these equations are just electroweak symmetry breaking (EWSB) conditions which are now solved for $m_1$ and $m_2$. So from the above, we see that $m_1$ and $m_2$ can now be expressed in terms of $\mu$ and $M_A$, which tells us that NUHM2 has the following free parameters:
\begin{equation}
m_0,~m_{1/2},~A_0,~\tan \beta,~\mu,~M_A,
\end{equation}
where the trilinear soft breaking term $A_0$ is not to be confused with the non-renormalizable term in inflationary scalar potential. 

In what follows, we want to find the regions of NUHM2 which are compatible with the allowed mass range for the Higgs boson and the observed dark matter abundance. We will use two methods. One consists in identifying benchmark points which will satisfy all these requirements, while the other method is more systematic and is based on a Markov Chain Monte Carlo (MCMC) scan of the NUHM2 parameter space.


\subsection{Identifying benchmark points for Neutralino dark matter}

To find interesting benchmark points, we use the \texttt{micrOMEGAs} code \cite{Belanger}, coupled to the Softsusy spectrum calculator \cite{Allanach} and impose the following requirements:

\begin{itemize}

\item The LSP must be a neutralino, 

\item The relic density of the neutralino must be compatible with the measured dark matter abundance by the WMAP experiment $0.1088<\Omega_{DM} h^2<0.1158$~\cite{WMAP}, 

\item The LEP2 bound on the mass of chargino must be satisfied. It is given by $m_{\chi^+_1}>103.5$ GeV \cite{chargino},

\item The mass of the lightest Higgs must be within the range that is not excluded yet at the LHC, i.e.~$[115.5, 127]$~GeV~\cite{ATLAS:2012ae,Chatrchyan:2012tx} and more precisely equal to either $m_h=119$ or  $m_h=125$ GeV. 
\end{itemize}

We scan the parameter space over the following range $0\leq m_0\leq 3000$ GeV and $0\leq m_{1/2}\leq 2000$~GeV (except for cases where we require the Higgs mass to be about $125$ GeV Higgs as this pushes the upper bound on $m_{1/2}$ to $m_{1/2}=4000$ GeV) and choose specific values of $\mu$ (the Higgs mixing parameter), $\tan\beta$  (the ratio of the VEVs for the $up$ and $down$ type fields) and $M_A$ (the mass of the pseudoscalar Higgs). 

We set the mass of the top quark to the Tevatron value, i.e. $m_t=173.2$ GeV \cite{TEV} and use the latest values of the following branching ratios $BR(B_s\rightarrow\mu^+\mu^-)<4.5\times10^{-9}$ \cite{Aaij:2012ac} and $BR(b\rightarrow s\gamma)=(3.55\pm0.26)\times10^{-4}$ \cite{Asner:2010qj}. 

None of the scenarios that we find below can explain the measured value of the anomalous magnetic moment of the muon $(g-2)_\mu$; the additional contributions in this model are indeed too small \cite{Baer:2011ab}. In what follows, we will assume that as long as the contribution of a given scenario is not greater than the measured value, the configuration is valid.

The same observation and assumption are made when we consider the branching ratio $BR(B^+ \rightarrow \tau^+ \bar \nu_\tau)$, knowing the latest best average $BR(B^+ \rightarrow \tau^+ \bar \nu_\tau)=(1.67\pm0.39)\times10^{-4}$ \cite{Asner:2010qj} 

\begin{figure}[t]
\centering
\includegraphics[width=1.05\linewidth]{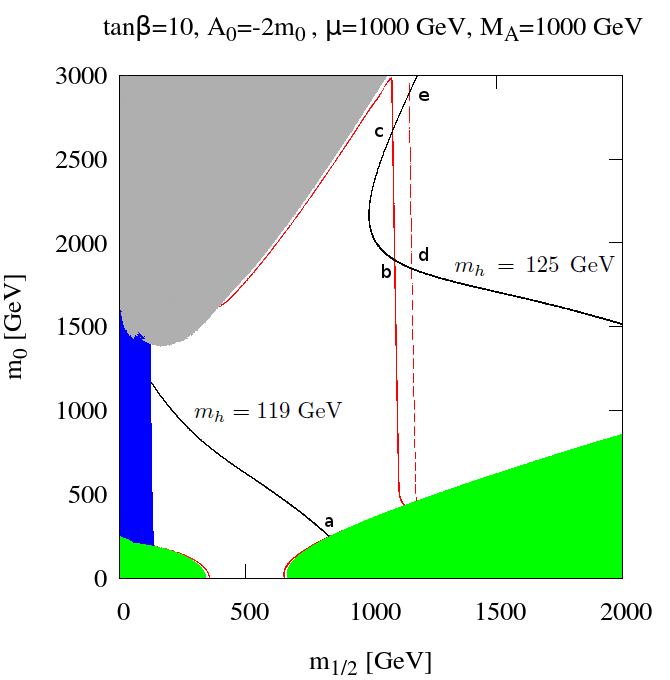}
\caption{$(m_0,m_{1/2})$ plane for the NUHM2 model. We explore specific configuration for $\tan\beta=10$, $A_0=-2m_0$, $\mu=1000$ GeV, $M_A=1000$ GeV. Red strips are where $0.1088<\Omega_{DM} h^2<0.1158$. Black lines show the two Higgs mass bounds.}
\label{fig1}
\end{figure}

Our first results are summarised in Fig.~(\ref{fig1}). 
In this figure the regions of the parameter space where the neutralino relic density is in agreement with the WMAP observations is represented by a red strip. The regions where the LSP is not a neutralino but a stau are coloured in green and the region excluded by the LEP2 limits on the chargino mass are represented in blue. The grey region corresponds to non physical configurations (in particular, we find that the stop is tachyonic in most of this region).

Since we are looking for points which satisfy both the  
Higgs and dark matter constraints, we define benchmark scenarios as 
the points which lie at the intersection between the red line representing the relic density and the two black lines corresponding respectively to a Higgs mass of $m_h=119$ and $m_h=125$ GeV. 

Our first conclusion for this choice of parameters is that 
 it is  hard to accommodate the correct LSP relic density with a Higgs mass of $m_h=119$ GeV. There is only a small overlap when $m_0=248$ GeV and $m_{1/2}= 834$ GeV (denoted by 'a') corresponding to 
$m_{\chi_1^0} = 351$ GeV and neutralino-stau co-annihilations.  
Indeed, below $m_{1/2}\approx 830$ GeV and for $m_0 < 500$ GeV, the 
stau and neutralino are almost mass degenerated. Hence the 
neutralino relic density mostly relies on neutralino-stau co-annihilations. For heavier neutralinos, both the coannihilation and annihilation rates decrease. As a result the LSP relic density becomes higher than the observed value.

Unlike the case for $m_h=119$ GeV, we find configurations intersecting the relic density and the $m_h=125$ GeV line. This assumes however that  $m_0 \geq 500$ GeV and $m_{1/2}\geq 800$ GeV. In this region, the correct LSP relic density is achieved through CP-odd Higgs s-channel self-annihilations. To explain the observed abundance, the neutralino mass must be close to (but not exactly on) the resonance region. This leads to the relation $m_{\chi_1^0}\approx M_A/2$ and thus implies that the neutralino mass is about $m_{\chi_1^0}\approx 500$ GeV for  $M_A=1000$ GeV. This region is actually referred to as the {\it funnel region}.  Between the two red strips, the relic density falls below the observed dark matter abundance because the annihilation process becomes resonant and reduces the relic density too much. In total, we thus identify four benchmark points when $m_h=125$ GeV. They are given by $m_0 = 1897,~ 2668,~ 1847,~ 2897$~GeV with $m_{1/2}\approx 1100$~GeV and correspond to 
the benchmark points ~'b',~'c',~'d',~'e'.

 
\begin{figure}[t]
\centering
\subfigure[]{\includegraphics[width=0.95\linewidth]{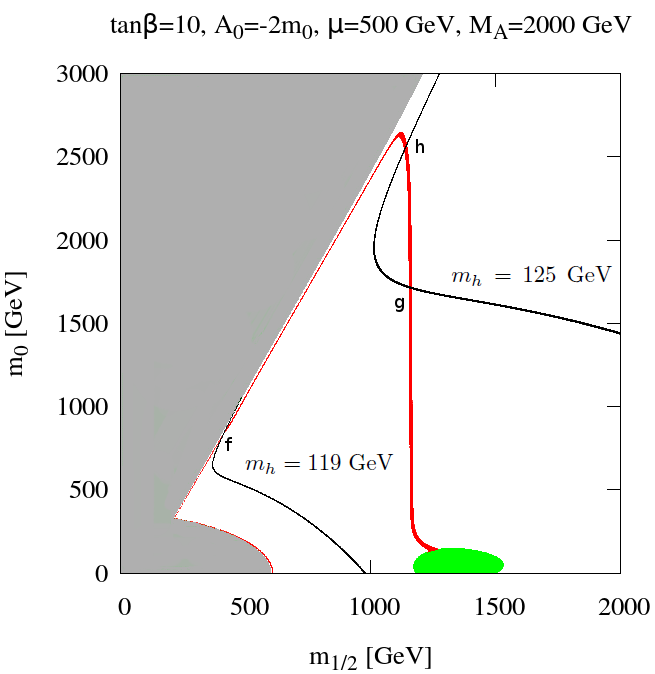}}\\
\subfigure[]{\includegraphics[width=0.95\linewidth]{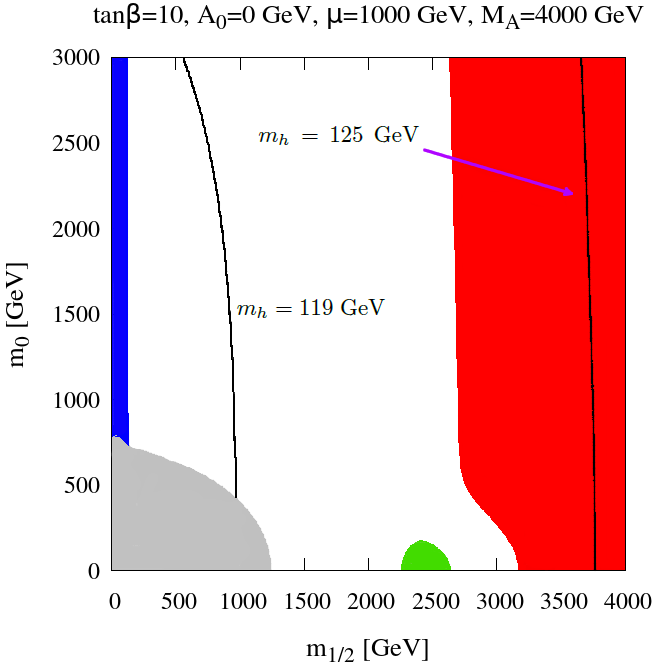}}
\caption{$(m_0,m_{1/2})$ plane for NUHM2: Panel (a) is for $A_0=-2m_0$, $\mu=500$ GeV, $M_A=2000$ GeV. Panel (b) is for $\mu=1000$ GeV, $A_0=0$ GeV and $M_A=4000$ GeV. Both panels share $\tan\beta=10$. The red regions show $0.1088<\Omega_{DM} h^2<0.1158$.}
\label{fig2}
\end{figure}


The situation is a bit different in Fig.~(\ref{fig2}) where $M_A$ is  larger. In panel (a) we find only one intersecting point for $m_h=119$ GeV which is given by $m_{1/2}=411$ GeV and $m_0=837$ GeV (benchmark point 'f'). In this region, the neutralino is mostly a bino.  Hence we expect its mass to be roughly equal to $M_1$, which is  related to $m_{1/2}$ via $M_1\approx 0.42 m_{1/2}$ at any RG scale. Therefore, this point corresponds to $M_1\approx m_{\chi_1^0} = 169$ GeV and the relic density is achieved through neutralino-stop coannihilation. 
Indeed the upper grey region just above this point, corresponding to $m_0 \geq 400$ GeV, denotes to a tachyonic stop. The latter becomes the NLSP at the edge of this region, thus leading to an acceptable neutralino relic density. Such a point is nevertheless very likely to be excluded by recent LHC searches \cite{Aad:2012cz}, even though the precise limit on the stop mass depends on the model that is considered and  Ref.~\cite{Aad:2012cz} assumed a gauge mediated scenario.

For $m_h = 125$ GeV, we can define two more benchmark points. 
Both have a large scalar mass $m_0$, namely $m_0=1715$ GeV (benchmark point 'g') and $m_0=2556$ GeV (benchmark point 'h') respectively with $m_{1/2}=1150$ GeV. Using the above relation between $M_1$ and $m_{1/2}$, we find that $M_1$ is larger than $\mu$, which implies an increasing fraction of Higgsino component in the neutralino~\cite{Ellis1}. 
Such a composition favours annihilation channels such as $\chi_1^0 \chi_1^0 \rightarrow W^+W^-,~Z Z,~Z h$ as well as neutralino-chargino co-annihilation and thus explains the vertical red strip in Fig.~\ref{fig2}(a). Since the mass of Higgsino-like neutralinos is primary sensitive to the Higgs mass mixing parameter $\mu$, we find that $m_{\chi_1^0}\approx 460$ GeV for both benchmark points, since they share the same $\mu$ and $m_{1/2}$.

In panel (b) of Fig.~(\ref{fig2}), we do not find any configuration compatible with both the 119 GeV Higgs and the observed dark matter abundance for the selected values of the $\tan \beta,\mu, M_A, A_0$ parameters. However, we note that, within explored range of $m_0$, any scalar masses at the GUT scale is compatible with $m_h=125$ GeV and the observed dark matter abundance. This leads to several possible benchmark points. The reason is that in this region the neutralino is very Higgsino-like. Therefore there is a large degeneracy between the two lightest neutralinos and the lightest chargino; hence the t-channel exchange process and the neutralino-neutralino and neutralino-chargino coannihilation mechanism contribute significantly to the LSP relic density \footnote{In the configurations of the parameter space leading to degenerate two lightest neutralinos and lightest chargino it is possible to embed the inflaton in a model where inflation is driven by the MSSM Higgses using the superpotential 
form: $W\supset\frac{\lambda_n}{n}\frac{(\bf{H_1\cdot H_
2})^n}{M_P^{2n-3}}$ \cite{
Chatterjee:2011qr}. In this paper we will not discuss this model.}. As the gaugino contribution becomes negligible,  $m_{\chi_1^0}$ becomes mostly sensitive to the $\mu$-parameter. As a result the LSP is slightly heavier than 1 TeV.

A summary of these benchmark points with their predictions for important observables is given in Table~(\ref{tab:summary}).

\begin{table*}
\begin{center}
\begin{tabular*}{.964\textwidth}{|c|c|c|c|c|c|c|c|c|c|c|c|}
\hline 
Fig./  &$(m_0;m_{1/2})$ & $\Omega h^2$ & $m_h$ & Dominant & $BR(B_s\rightarrow\mu^+\mu^-)$ & $BR(b\rightarrow s\gamma)$ & $g_\mu-2$ & $m_{\chi_1^0}$ &Channel&$m_{\phi_{\widetilde u\widetilde d\widetilde d}}$&$m_{\phi_{\widetilde L\widetilde L\widetilde e}}$ \\ 
Label &(GeV)  &   & (GeV) & component(s) & ($10^{-9}$) & ($10^{-4}$) & ($10^{-11}$) & (GeV) & & (GeV) & (GeV)\\
\hline 
\hline 
      \ref{fig1} 'a' &(248;834) & 0.111 & 119 & $\widetilde{B}$ & 3.085 & 3.348 & 30.7 & 351 & $\tilde{\tau}$ & 955 & 438 \\ \hline
      \ref{fig1} 'b' &(1897;1093) & 0.112 & 125 & $\widetilde{B}$ & 3.113 & 3.305 & 6.07 & 473 & A & 2249 & 1955 \\ \hline
      \ref{fig1} 'c' &(2668;1085) & 0.111 & 125 & $\widetilde{B}$ & 3.130 & 3.241 & 3.96 & 473 & A & 2925 & 2709 \\ \hline
      \ref{fig1} 'd' &(1847;1161) & 0.113 & 125 & $\widetilde{B}$ & 3.108 & 3.330 & 6.05 & 503 & A & 2249 & 1914 \\ \hline
      \ref{fig1} 'e' &(2897;1152) & 0.112 & 125 & $\widetilde{B}$ & 3.130 & 3.252 & 3.46 & 503 & A & 3165 & 2939 \\ \hline
      \ref{fig2} (a) 'f' &(837;411) & 0.109 & 119 & $\widetilde{B}$ & 3.090 & 1.731 & 36.1 & 169 & $\widetilde{t}$ & 952 & 855 \\ \hline
      \ref{fig2} (a) 'g' &(1715;1158) & 0.111 & 125 & 0.69$\widetilde{B}$+0.31$\widetilde{H}$ & 3.076 & 3.092 & 7.82 & 465 & $\chi^{+,0}$ & 2140 & 1787 \\ \hline
      \ref{fig2} (a) 'h' &(2556;1140) & 0.110 & 125 & 0.71$\widetilde{B}$+0.29$\widetilde{H}$ & 3.080 & 2.921 & 4.57 & 462 & $\chi^{+,0}$ & 2850 & 2603 \\ \hline
\end{tabular*}
\caption{\label{tab:summary}Benchmark points considered in this study and associated predictions for important observables. The Figures  which they are associated to and the dominant mechanism ($\tilde{\tau}, \tilde{t}$ coannihilations, $\chi^{+,0}$ exchange, A-pole) for the relic density calculations are specified in the last two columns of the table. The mass of the inflaton is at low scale.}
\end{center}
\end{table*}


\subsection{A broader scan of the parameter space}

In the previous subsection we have identified a set of parameters for which the Higgs mass coincided with either $m_h=$119 or $m_h=$125 GeV, and simultaneously lead to a dark matter relic density compatible with WMAP observations~\cite{WMAP}.
We now want to check whether the predictions associated with these benchmark points are generic or not.

We thus perform a more general scan of the NUHM2 parameter space. We now want to identify the regions of the parameter space which lead to a Higgs mass within $[115.5, 127]$ GeV and a neutralino relic density within the WMAP measurements, namely 
$\Omega_{DM} h^2 \in [0.1088,0.1158]$ using WMAP 7-year $+$ BAO $+ H_0$ mean value\cite{WMAP}. 

For this purpose, we use a MCMC coupled to the \texttt{micrOMEGAs} code along the lines described in \cite{Vasquez:2010ru}. The total likelihood function is computed for each point chosen in the parameter space and is the product of the likelihood functions associated with each observable. 
Since we are not interested in characterising how statistically relevant the points that we found are but want instead to determine the full range of configurations that are possible,  we will not account for the number of occurrence of a given scenario. The drawback of such a method is that 
we cannot determine how likely a region of the parameter space is with respect to the other parts. The advantage is that very small (fine-tuned) configurations are kept in the analysis.

For the Higgs mass and relic density, we define the likelihood as  a function $\mathcal{L}_1$ which decays exponentially at the edges of   the $[x_{min}, x_{max}]$ range, according to 
\begin{align}
\mathcal{L}_1(x, x_{min}, x_{max}, \sigma)  = & \: e^{-\frac{\left( x - x_{min}\right)^2}{2\sigma ^2}} \: \textrm{if} \: x < x_{min}, \nonumber \\
= & \: e^{-\frac{\left( x - x_{max}\right)^2}{2\sigma ^2}} \: \textrm{if} \: x > x_{max}\nonumber \\
= & \: 1 \: \textrm{for} \: x \in [x_{min}, x_{max}].
\end{align}
with $\sigma$ a variance corresponding to the width of the $[x_{min}, x_{max}]$ range and $x$ the observable which corresponds in that case to either the Higgs mass or the LSP relic density.   

For all the other observables, we will use two types of likelihood.
\begin{itemize}
\item For an observable with a preferred value $\mu$ and error $\sigma$, we use a Gaussian distribution $\mathcal{L}_2$ :
\begin{equation}
\mathcal{L}_2(x, \mu, \sigma) = e^{-\frac{\left(x-\mu\right)^2}{2\sigma ^2}}.
\end{equation}
\item For an observable with a lower or upper bound (set experimentally),  we will take the function $\mathcal{L}_3$ with a positive or negative variance $\sigma$ :
\begin{equation}
\mathcal{L}_3(x, \mu, \sigma) = \frac{1}{1+e^{-\frac{x-\mu}{\sigma}}}.
\end{equation}
\end{itemize}
We assume flat priors for all the parameters considered in this paper, and immediately reject configurations where at least one of the parameters fall outside the specified range. Points for which the calculation of the SUSY spectrum fail (i.e. when there is no electroweak symmetry breaking or there is the presence of tachyonic particles) or the neutralino is not the LSP are also immediately rejected.  At last, we do not implement the limits on sparticle masses from the LHC, since the squark masses that we consider are above the present limits. LEP limits on sleptons (and squarks) are nevertheless taken into account in \texttt{micrOMEGAs}.

\begin{table*}[!htb]
\begin{tabular*}{0.8854\textwidth}{ | c | c | c | c | }
\hline Constraint & Value/Range & Tolerance & likelihood \\ 
\hline \hline
       $m_h$ (GeV) \cite{ATLAS:2012ae,Chatrchyan:2012tx} & [115.5, 127] & 1 & $\mathcal{L}_1(m_h, 115.5, 127, 1)$ \\ \hline 
       $\Omega_{\chi^0_1} h^2$ \cite{WMAP} & [0.1088, 0.1158] & 0.0035 & $\mathcal{L}_1(\Omega_{\chi^0_1} h^2, 0.1088, 0.1158, 0.0035)$ \\ 
       Relaxing constraint on $\Omega_{\chi^0_1} h^2$ & [0.01123, 0.1123] & 0.0035 & $\mathcal{L}_1(\Omega_{\chi^0_1} h^2, 0.01123, 0.1123, 0.0035)$ \\\hline
       $BR(b \rightarrow s\gamma)$ $\times$ $10^{4}$ \cite{Asner:2010qj,Misiak:2006zs} & 3.55 & exp : 0.24, 0.09 & $\mathcal{L}_2(10^{4} BR(b \rightarrow s \gamma), 3.55,$ \\ && th : 0.23 &  $\sqrt{0.24^2 + 0.09^2 + 0.23^2})$\\ \hline
       $(g_\mu-2)$ $\times$ $10^{10}$ \cite{Davier:2010nc} & 28.7 & 8 & $\mathcal{L}_3(10^{10} (g_\mu-2), 28.7, -8)$ \\ \hline 
       $BR(B_s \rightarrow \mu^+ \mu^-)$ $\times$ $10^{9}$\cite{Aaij:2012ac} & 4.5 & 0.045 & $\mathcal{L}_3(10^{9} Br(B_s \rightarrow \mu^+ \mu^-), 4.5, -0.045)$ \\ \hline        
       $\Delta \rho$ & 0.002 & 0.0001 & $\mathcal{L}_3(\Delta \rho, 0.002, -0.0001)$ \\ \hline 
       $R_{B^+ \rightarrow \tau^+ \bar \nu_\tau} (\frac{NUHM2}{SM})$ \cite{Charles:2011va} & 2.219 & 0.5 & $\mathcal{L}_3(R_{B^+ \rightarrow \tau^+ \bar \nu_\tau}, 2.219, -0.5)$ \\ \hline 
       $Z \rightarrow \chi^0_1 \chi^0_1$ (MeV) & 1.7 & 0.3 & $\mathcal{L}_3(Z \rightarrow \chi^0_1 \chi^0_1, 1.7, -0.3)$ \\ \hline 
       $\sigma_{e ^+ e ^- \rightarrow \chi^0_1 \chi^0_{2,3}}$ & 1 & 0.01 & $\mathcal{L}_3(\sigma_{e ^+ e ^- \rightarrow \chi^0_1 \chi^0_{2,3}} \times Br(\chi^0_{2,3} \rightarrow Z \chi^0_1),1, -0.01)$ \\ 
$\times Br(\chi^0_{2,3} \rightarrow Z \chi^0_1)$ (pb) \cite{Abbiendi:2003sc} &&& \\ 
\hline 
\end{tabular*}
\caption{\label{tab:constraints}Constraints imposed in the MCMC, from \cite{Nakamura:2010zzi} unless noted otherwise.}
\end{table*}


The known constraints that we impose from Particle Physics are summarised in Table.~(\ref{tab:constraints}) and the range that we consider for the different parameters are given in Table~(\ref{tab:range}).

\begin{table}[!htb]
\begin{center}
\begin{tabular}{|c|c|}\hline
Parameter & Range \\ \hline \hline
$m_{0}$ & ]0, 4] TeV\\ 
$m_{1/2}$ & ]0, 4] TeV\\ 
$A_{0}$ & [-6, 6] TeV\\ 
$\tan \beta$ & [2, 60]\\
$\mu$ & ]0, 3] TeV\\ 
$M_A$ & ]0, 4] TeV\\ \hline
\end{tabular}
\caption{Range chosen for the free parameters in the NUHM2 model. \label{tab:range}}
\end{center}
\end{table}
%

\begin{figure}[h]
\centering
\includegraphics[scale=0.23]{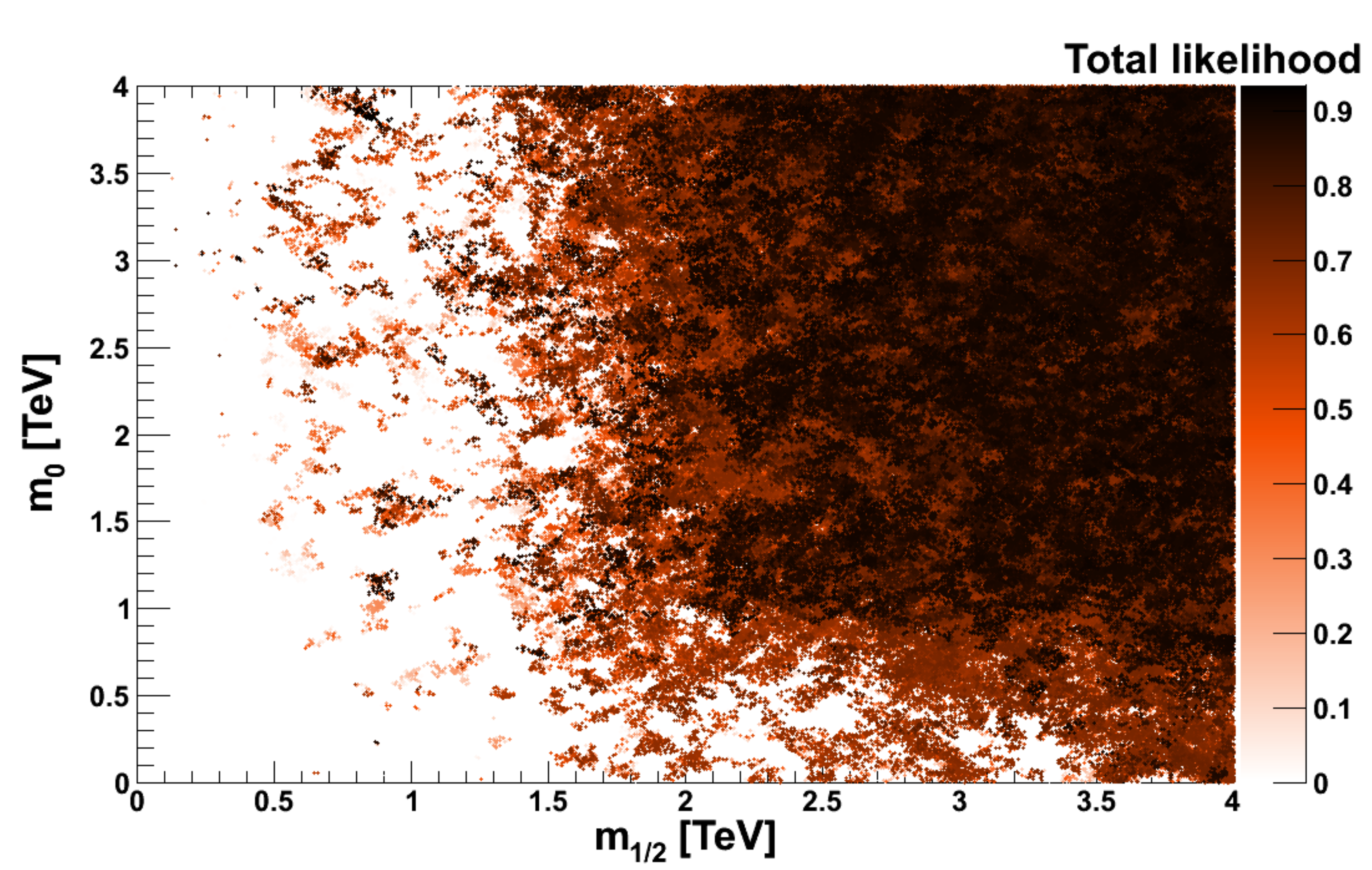}
\caption{Plot of the allowed parameter space in the $(m_0,m_{1/2})$ plane. We use the likelihood of the points as colour code. The darkest points have the highest likelihood. However they may not be statistically significant. }
\label{mo_mhalf_mcmc}
\end{figure}

In Fig.~(\ref{mo_mhalf_mcmc}), we see that most of the scenarios 
found by the MCMC involve TeV scale values of $m_0$ and $m_{1/2}$, but no real feature emerges from the plot.


\begin{figure}[h]
\centering
\includegraphics[scale=0.23]{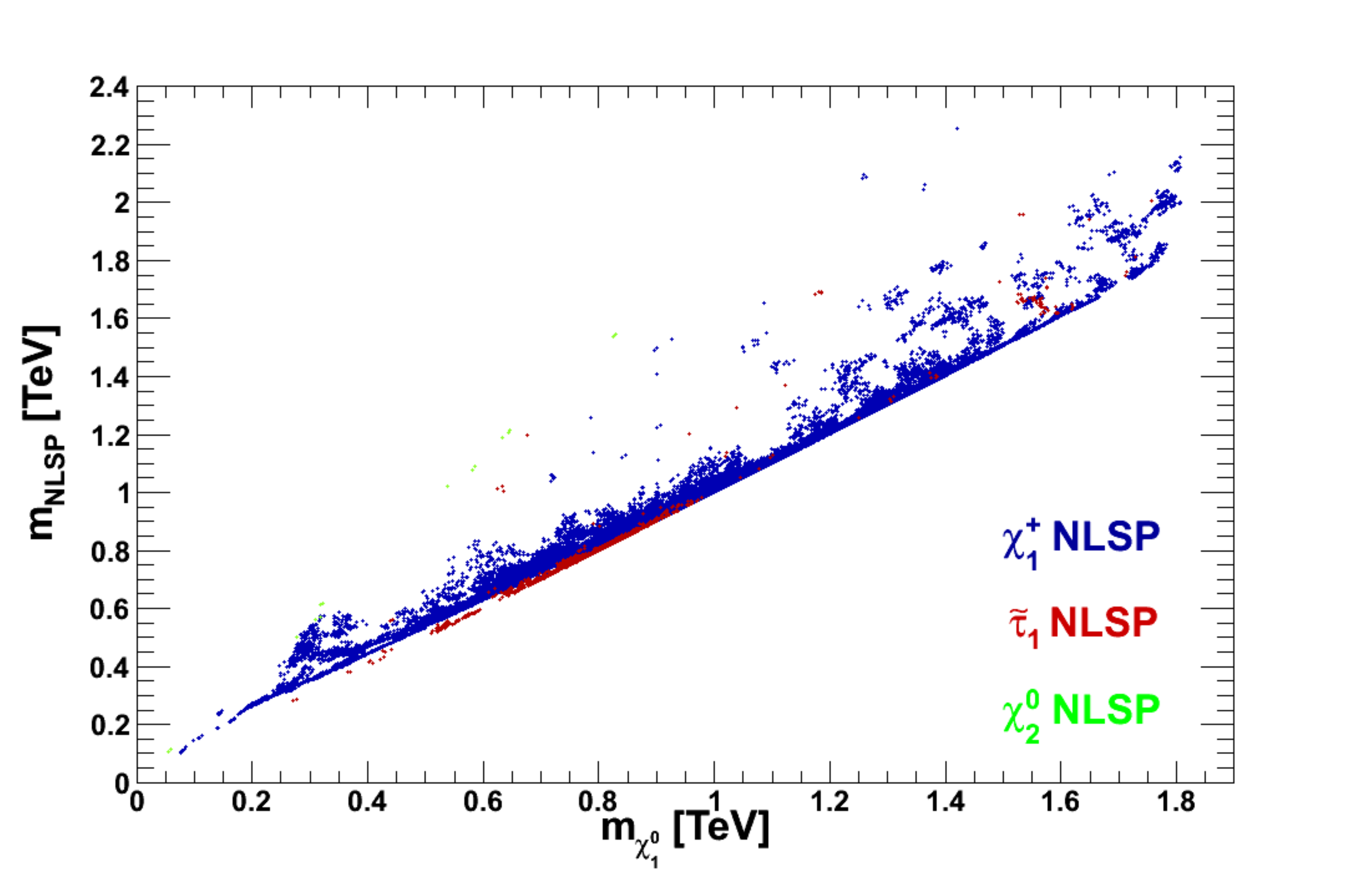}
\caption{Mass of the LSP vs the mass of the NLSP, depending on the nature of the NLSP. Only points with dominant likelihood were considered.}
\label{mLSP_mNLSP_mcmc}
\end{figure}


Nevertheless, as illustrated  in Fig.~(\ref{mLSP_mNLSP_mcmc}), there is a very strong correlation between the mass of the LSP and that of the NLSP, suggesting that the neutralino relic density either relies on the co-annihilation mechanism 
or a $t$-channel exchange of the NLSP (or both).  The NLSP is found to be mostly a chargino, a neutralino and a stau  as 
obtained for the benchmark points 'a','g','h'. The $A$-pole resonance corresponding to the benchmark points 'b','c','d','e' 
requires however a certain amount of fine tuning (precisely because it requires $m_{\chi^0_1} \simeq M_A/2$) and is not 
the most represented configuration found by the MCMC. 


\begin{figure}[h]
\centering
\includegraphics[scale=0.23]{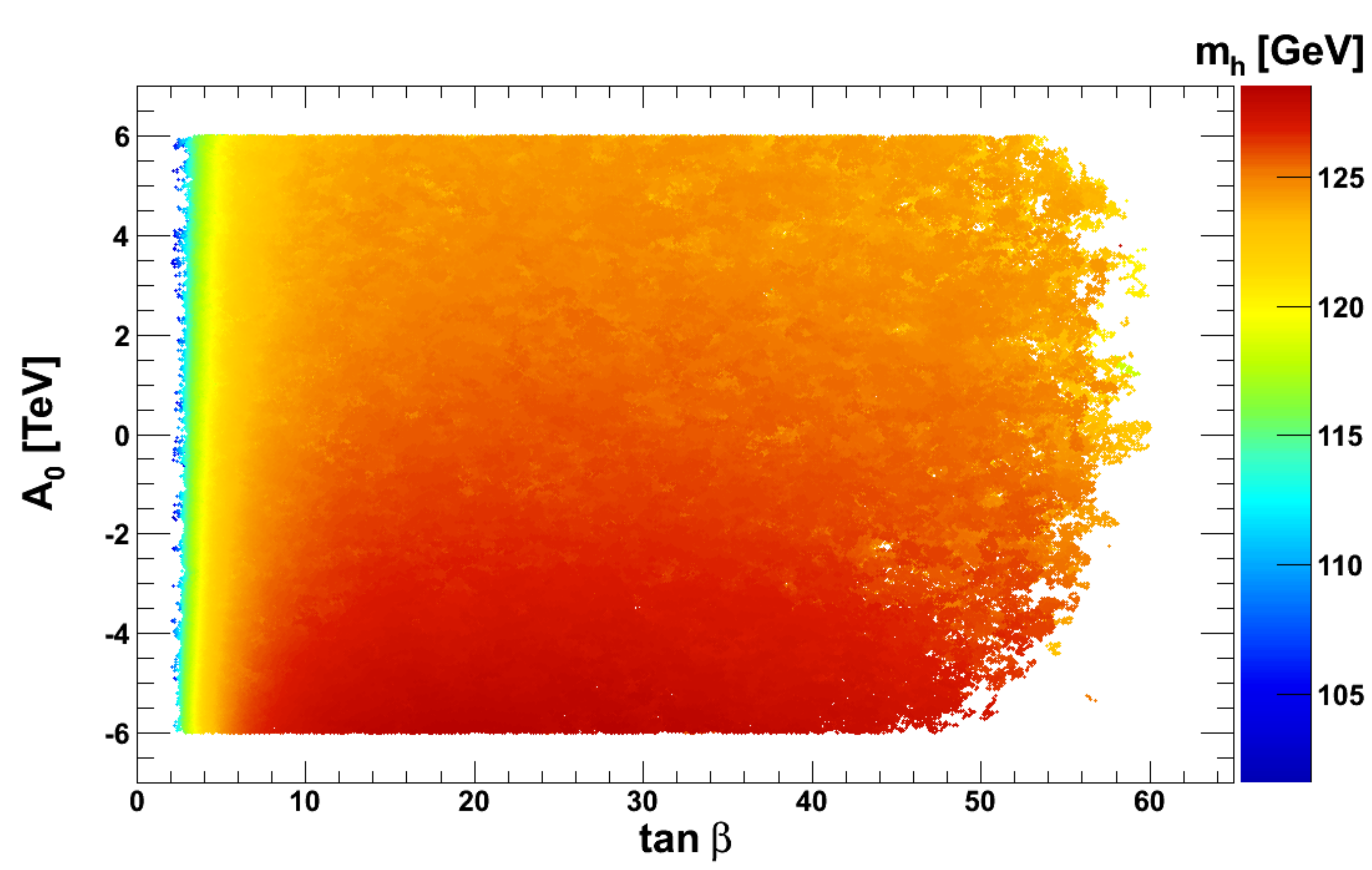}
\caption{Higgs mass in the $ (A_0,\tan \beta)$ plane. Light Higgs can be found whatever the value of the trilinear coupling $A_0$, provided that $\tan \beta$ is small.} 
\label{a0_vs_tb_vs_mh}
\end{figure}


The predominance of scenarios in which charginos are mass degenerated with neutralinos can be understood by inspecting 
Fig.~(\ref{a0_vs_tb_vs_mh}). For the configurations with $A_0 = -2 \ m_0$, the Higgs mass $m_h$ tends to exceed the upper experimental bound unless one decreases the value of $\tan \beta$~\footnote{Note that we didn't computed the amount of ElectroWeak fine-tuning in our NUHM2 scenarios. It was shown, for instance in \cite{Ellis:2007by}, that some NUHM2 benchmark points wherein $A_0 = 0$~TeV give non-negligible EW fine-tuning.}. 
.

For such configurations, the sparticle masses are generally too large for the sparticle-neutralino  co-annihilation channels to reduce the relic density significantly and both the neutralino and chargino have a significant Higgsino fraction, Fig.~\ref{neutralino_composition}. As a result, the possible channels to reduce the neutralino relic density 
either involve CP-odd Higgs portal annihilations or neutralino-chargino mass degeneracies.  

\begin{figure}[h]
\centering
\includegraphics[scale=0.22]{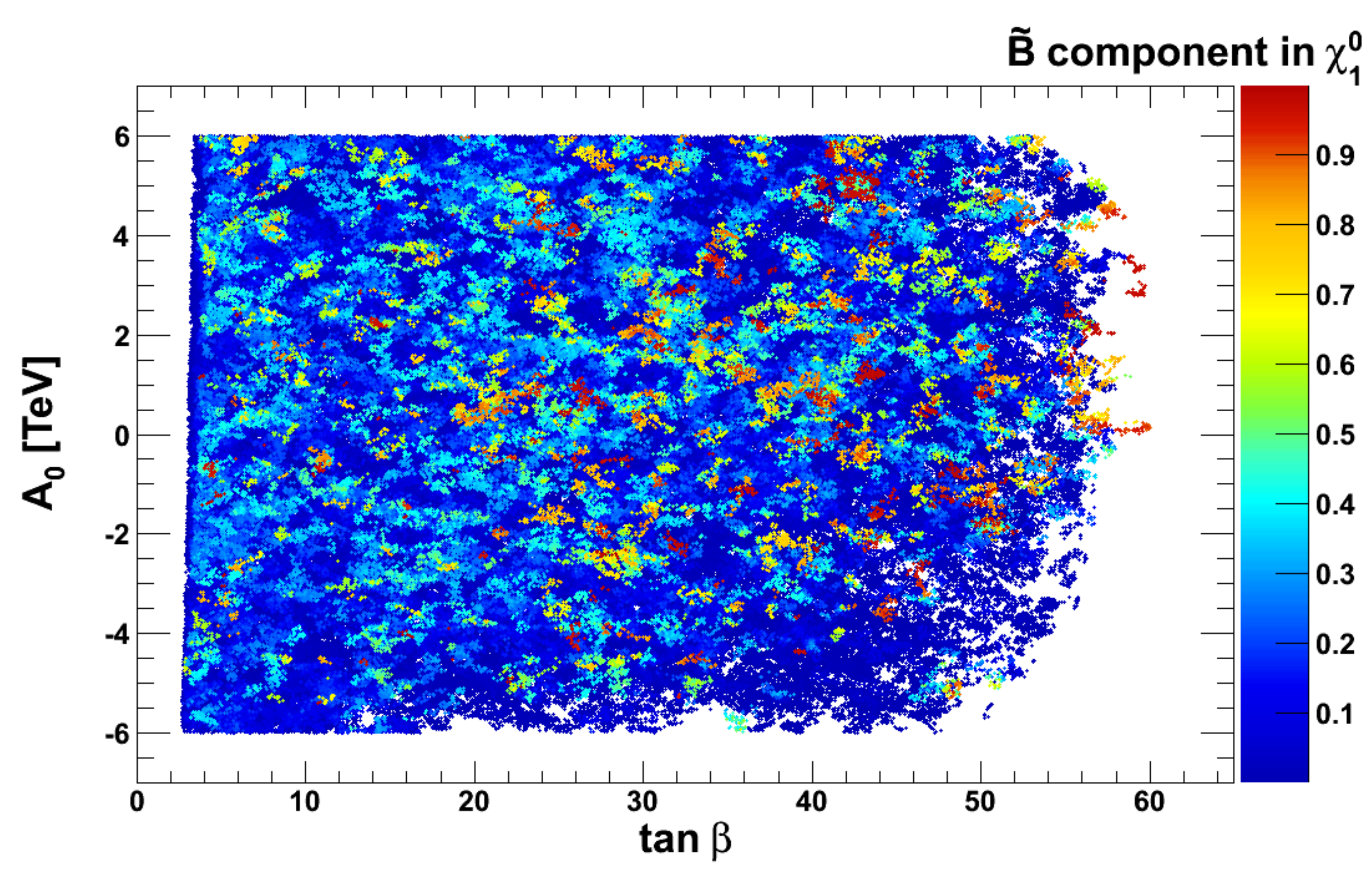}
\caption{Bino fraction in the $ (A_0,\tan \beta)$ plane.} 
\label{a0_vs_tb_vs_Bino}
\end{figure}

\begin{figure}[h]
\centering
\includegraphics[scale=0.22]{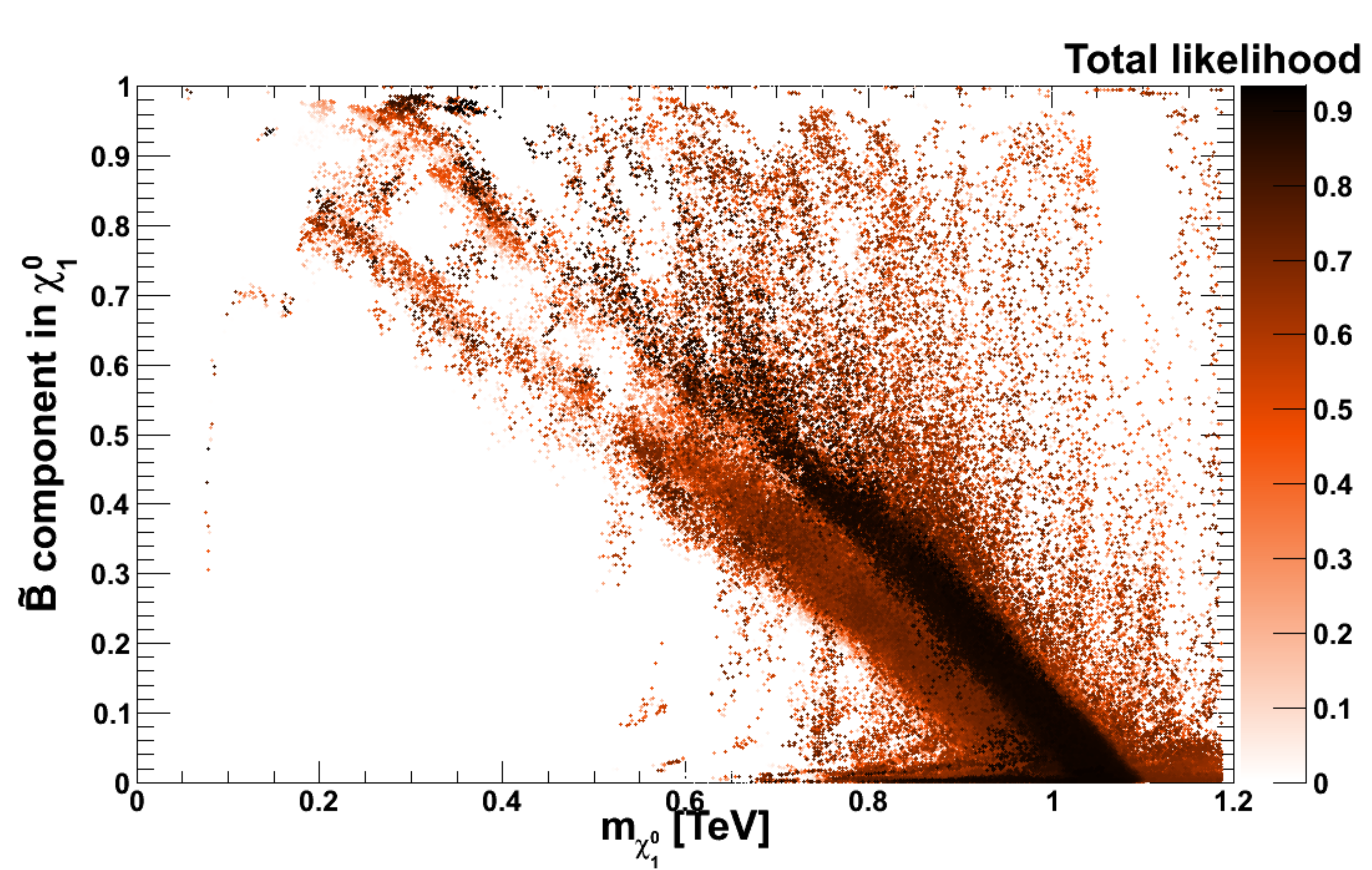}  \newline
\includegraphics[scale=0.22]{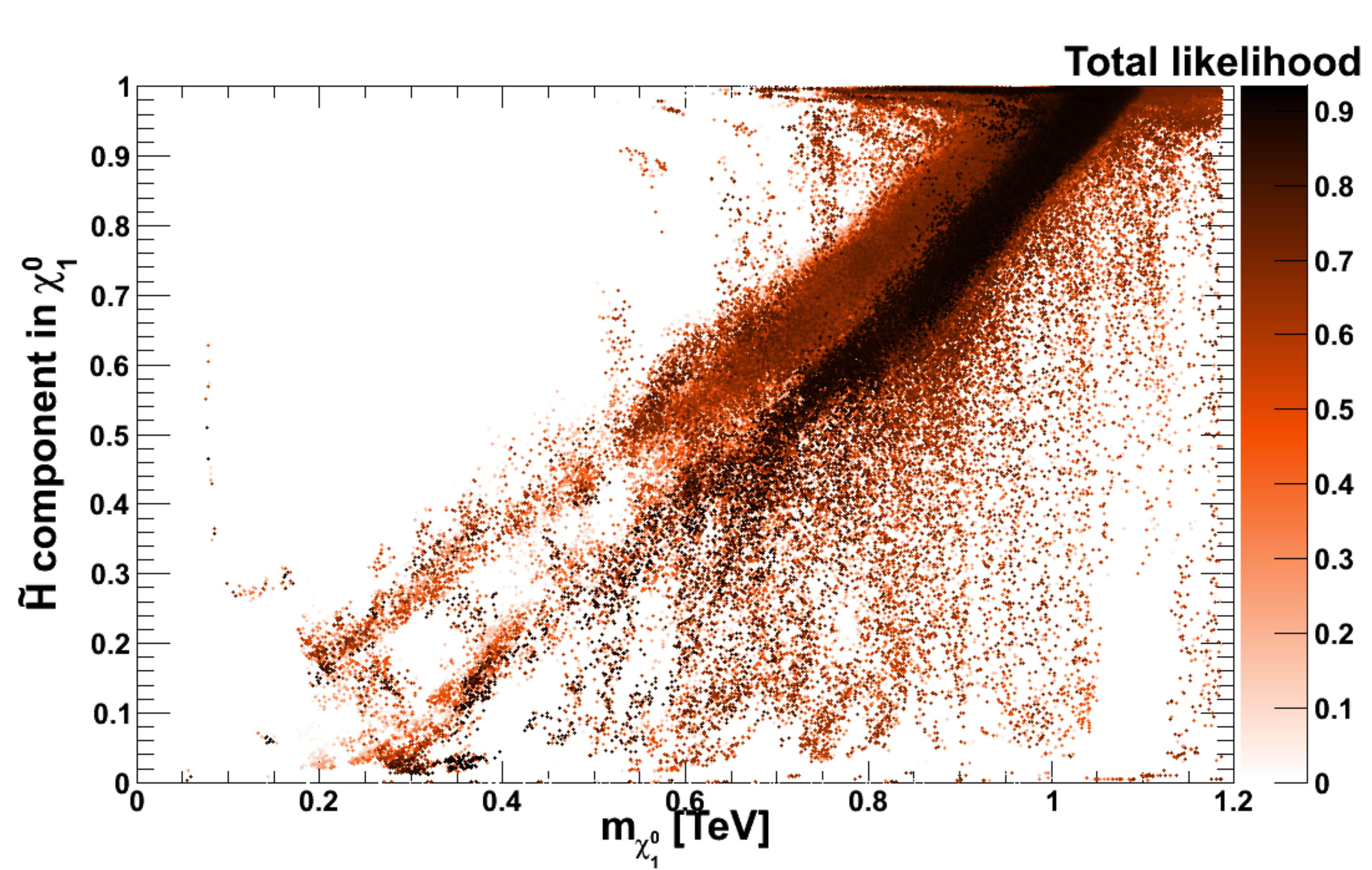}
\caption{Neutralino composition. Top panel shows the bino content vs the neutralino mass while the bottom panel shows the Higgsino fraction. The colour coding corresponds to the likelihood of these points.} 
\label{neutralino_composition}
\end{figure}

The exchange of a pseudoscalar Higgs is actually significant when $m_{\chi^0_1} \sim M_A/2$ (as found for the benchmark points 'b','c','d','e') but neutralino-chargino coannihilation or chargino t-channel exchange are dominant when the Higgsino fraction is very large. In fact, among the configurations with a non-negligible Higgsino fraction, 
the larger the bino fraction, the more favoured the A-pole since 
small neutralino couplings to the Higgs can be compensated by having $m_{\chi^0_1}$ closer to $M_A$. 
The distribution of points depending on their bino fraction is represented in the plane $(A_0,\tan \beta)$ in 
Fig.~(\ref{a0_vs_tb_vs_Bino}). Clearly scenarios with Bino-like neutralinos are under represented, illustrating how fine-tune they are.


Finally, we see from Fig.~(\ref{neutralino_composition}) that heavy neutralinos with a mass 
$m_{\chi}\geq 0.6$ TeV have a large Higgsino fraction, thus suggesting even more dominant   coannihilations with charginos (or annihilations through chargino exchange) and resonant annihilations via the pseudoscalar Higgs when the neutralino becomes fairly heavy.   
Interestingly though, for most these scenarios, the value of the $\mu$ parameter varies between $500$ GeV and $1.5$ TeV but the  values which correspond to the highest likelihood are about $\mu \simeq 1$~TeV, which is indeed consistent with a large Higgsino fraction.


\begin{figure}[h]
\centering
\includegraphics[width=1.0\linewidth]{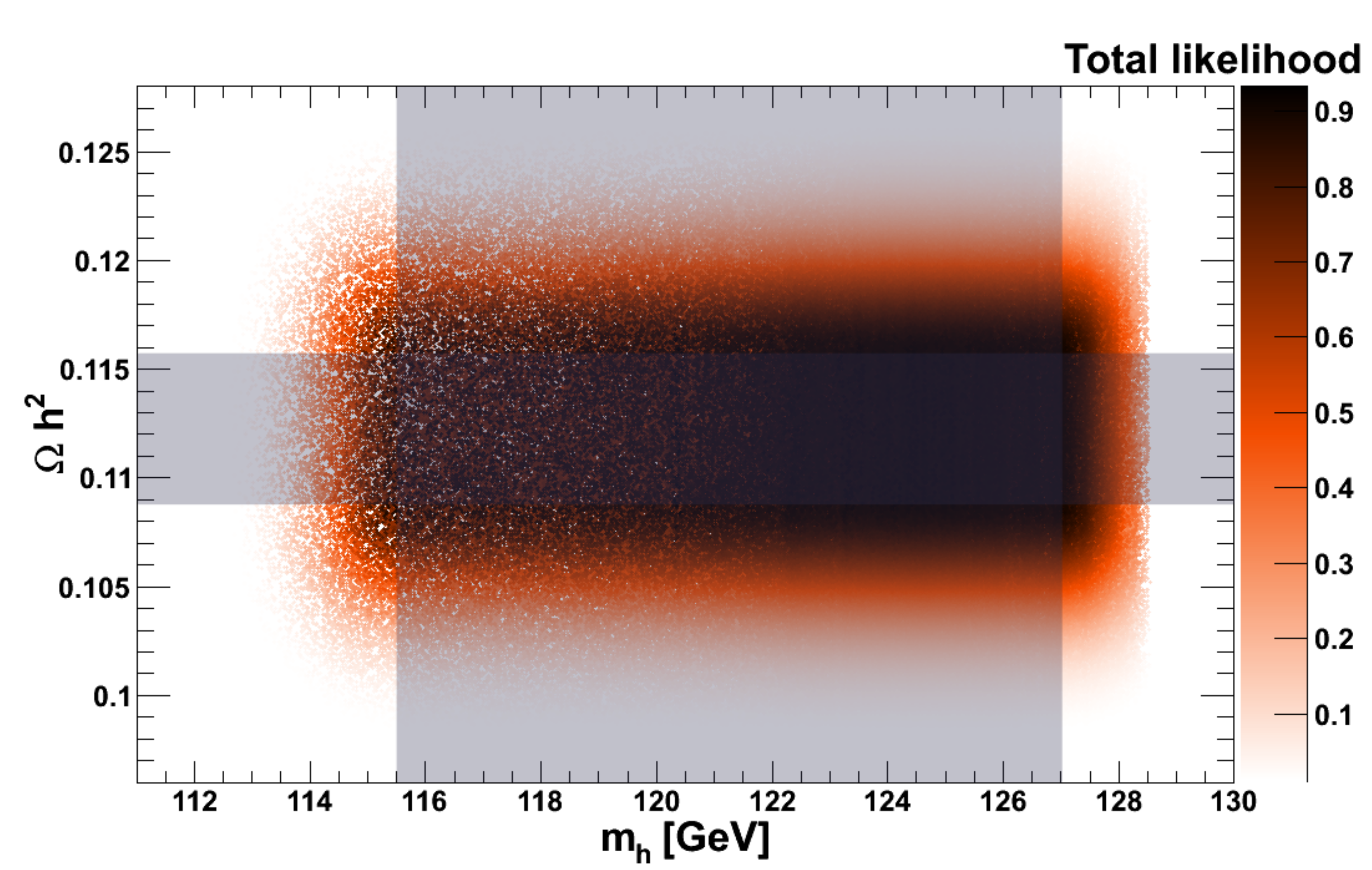}
\caption{Neutralino relic density vs the mass of the Higgs. The distribution of points show that any Higgs mass is associated to a high likelihood.}
\label{RDHiggs}
\end{figure}

We can now investigate the distribution of points which satisfy the 
constraints on the Higgs mass and the dark matter relic density, Fig.~(\ref{RDHiggs}). The points with high
likelihood are 'smoothly' distributed within the observed relic density and Higgs mass range. 


Relaxing the constraint on the dark matter relic density and allowing  neutralinos to constitute only a fraction of the total dark matter energy density does not change the above features. 
The main effect in fact is to allow lower values of $\mu$ 
and reduce the mass degeneracy between the LSP and NLSP. However these degeneracies are still present and larger values of the Higgs mass still give a higher likelihood.


\section{Indirect detection of the inflaton at LHC}
\label{IM}


\begin{figure}[t]
\centering
\subfigure[]{\includegraphics[width=0.85\linewidth]{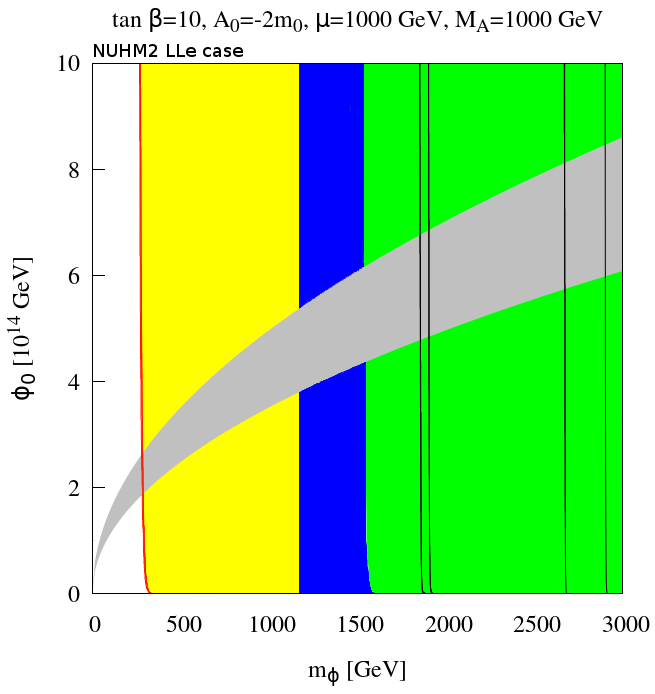}}  
\subfigure[]{\includegraphics[width=0.85\linewidth]{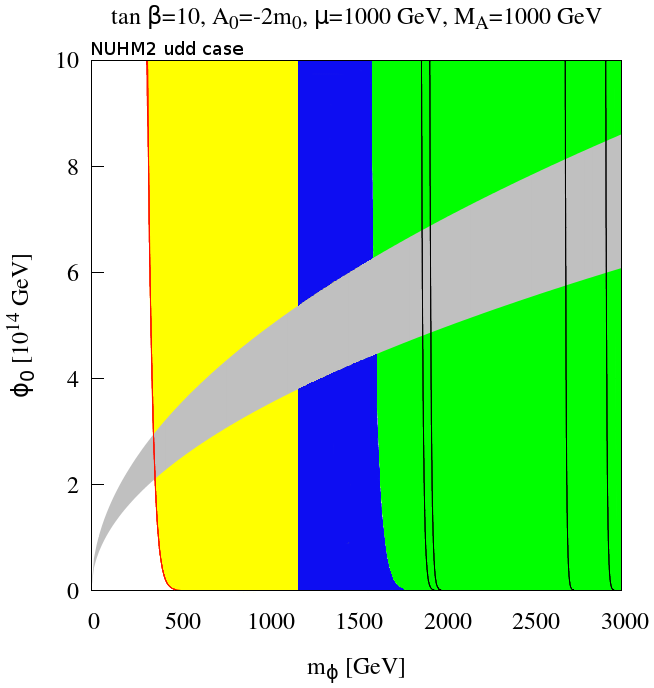}}
\caption{($\phi_0$,$m_\phi$) plane for $\widetilde{L}\widetilde{L}\widetilde{e}$ and $\widetilde{u}\widetilde{d}\widetilde{d}$ flat direction inflatons respectively, where $\tan\beta=10$, $A_0=-2m_0$, 
$\mu=1000$ GeV, and $M_A=1000$ GeV. The yellow region corresponds to the Higgs mass of $119$~GeV and the green region  corresponds to that of the $125$ GeV. Blue region is the same as in Fig.~(\ref{fig1}) -- excluded because of the mass bounds on chargino. Red and Black strips show where the dark matter abundance match
within $0.1088<\Omega_{DM} h^2<0.1158$. }
\label{fig3}
\end{figure}


In the previous section we have verified the validity of our benchmark 
points and could measure how fine-tune they are with respect to other configurations. In particular, 
we have seen that scenarios with large scalar masses $m_0$ require 
small values of $\tan \beta$ in order to not exceed the upper limit on the Higgs mass and lead to scenarios in which the neutralino has generally a non-negligible Higgsino fraction. 

We can now determine the inflation energy scale and mass of the inflaton 
for these benchmark points. We will follow a similar approach as in Ref.~\cite{Allahverdi:2010zp} in order to estimate the inflaton mass which is compatible with the temperature anisotropy of the CMB data.


\subsection{Inflaton mass for benchmark points}

In Fig.~(\ref{fig3}) we have mapped the regions of Fig.~(\ref{fig1}) onto $(\phi_0,m_\phi)$ plane. Panel (a) is for the $\widetilde{L}\widetilde{L}\widetilde{e}$ case and panel (b) is for the $\widetilde{u}\widetilde{d}\widetilde{d}$ case. The yellow region is where the Higgs mass is equal to $119$ GeV, and the green one is for $m_h = 125$ GeV. The blue area correspond to the LEP2 bound on the chargino mass. Vertical lines represent the inflaton mass for which the Higgs mass intersects with the WMAP relic density measurements. We used the red color for $m_h = 119$ GeV, and black for $m_h=125$ GeV. The grey shaded region shows where the NUHM2 inflation can explain the CMB observations. From these figures we see that if we 
want consistent description of $119$ GeV Higgs mass with the dark matter density measurements, inflation must happen in a range roughly given by: 
$\phi_0 \approx(1.8-2.6)\times10^{14}$ GeV for $\widetilde{L}\widetilde{L}\widetilde{e}$ and $\phi_0 \approx(2.1-2.8)\times10^{14}$ GeV for $\widetilde{u}\widetilde{d}\widetilde{d}$ inflaton candidates, corresponding to a mass of $m_\phi\approx300$ GeV and $m_\phi\approx380$ GeV respectively.


\begin{figure}[t]
\centering
\subfigure[]{\includegraphics[width=0.85\linewidth]{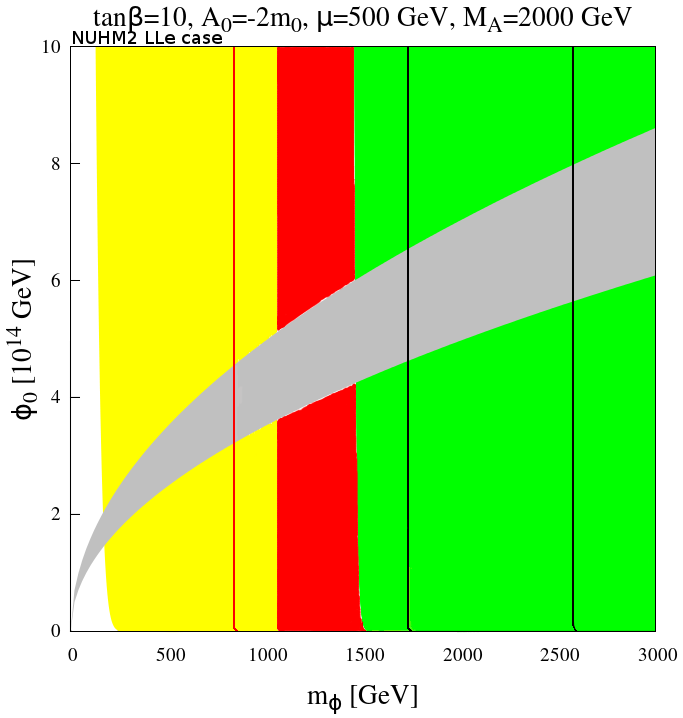}}\\

\subfigure[]{\includegraphics[width=0.9\linewidth]{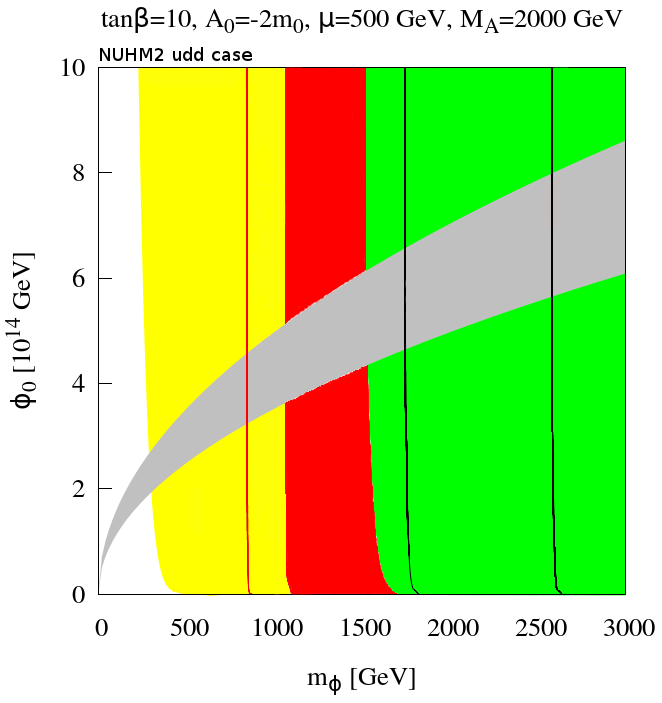}}
\caption{($\phi_0$,$m_\phi$) plane for $\widetilde{L}\widetilde{L}\widetilde{e}$ and $\widetilde{u}\widetilde{d}\widetilde{d}$ flat direction inflatons, where $\tan\beta=10$, $A_0=-2m_0$, the 
$\mu=500$ GeV, and $M_A=2000$ GeV. The yellow region corresponds to the Higgs mass of $119$~GeV and the green region  corresponds to that of a $125$ GeV Higgs. The red and black strips show where the dark matter abundance fall within $0.1088<\Omega_{DM} h^2<0.1158$.}
\label{fig4}
\end{figure}


In the case of $m_h=125$ GeV Higgs, inflation should happen around $\phi_0\approx(4.8-6.8)\times10^{14}$~GeV for the 'b' and 'd' benchmark points, see Table.~\ref{tab:summary}, yelding $m_\phi\approx 1900$ GeV for the $\widetilde{L}\widetilde{L}\widetilde{e}$  scenario and a slightly heavier $\widetilde{u}\widetilde{d}\widetilde{d}$ candidate. Another two possibilities correspond to the bechmark points 'c' and 'e', see Table~\ref{tab:summary}. For 'c' point we have inflation happening in a range of $\phi_0\approx(5.7-8)\times10^{14}$ GeV with a mass of the inflaton being around $m_\phi\approx 2700$ GeV and similarly for 'd' we have range of $\phi_0\approx(6-8.1)\times10^{14}$ GeV with $m_\phi\approx 2950$ GeV. From cosmological point of view, the heavier Higgs boson is, the more of the parameter space for inflation which become compatible with the CMB observations we have. In general for the $\widetilde{u}\widetilde{d}\widetilde{d}$ inflaton, we have a larger running than in the $\widetilde{L}\widetilde{L}\
widetilde{e}$ case, essentially because of the running of $g_3$. However, it is hard to appreciate this running 
visibly by comparing Figs.~\ref{fig3} (a) and (b), because of the large range of $m_\phi$ we have  plotted.

In Fig.~\ref{fig4} we are mapping the points of Fig.~\ref{fig2} (a). Here the red shaded region corresponds to the allowed relic density, whereas the yellow and green regions 
are the same as in the previous discussion. The Higgs mass of $m_h = 119$ GeV again implies the lower scale for inflation -- i.e. $\phi_0\approx(3-4.2)\times10^{14}$~GeV, with an inflaton mass of around $m_\phi\approx 860$ GeV. However for a Higgs mass of $125$ GeV, we find two energy scales related to different $m_\phi$ values:   $\phi_0\approx(4.2-6.2)\times10^{14}$~GeV when $m_\phi\approx 1780$ GeV, and $\phi_0\approx(5-7.5)\times10^{14}$~GeV when $m_\phi\approx2550$ GeV. 
We obtain similar conclusions for Fig.~(\ref{fig4}) (b), where the $\widetilde{u}\widetilde{d}\widetilde{d}$ direction gives a slightly higher mass for the inflaton. 


\subsection{LHC predictions and Inflaton mass}

\begin{figure}[h]
\centering
\includegraphics[scale=0.23]{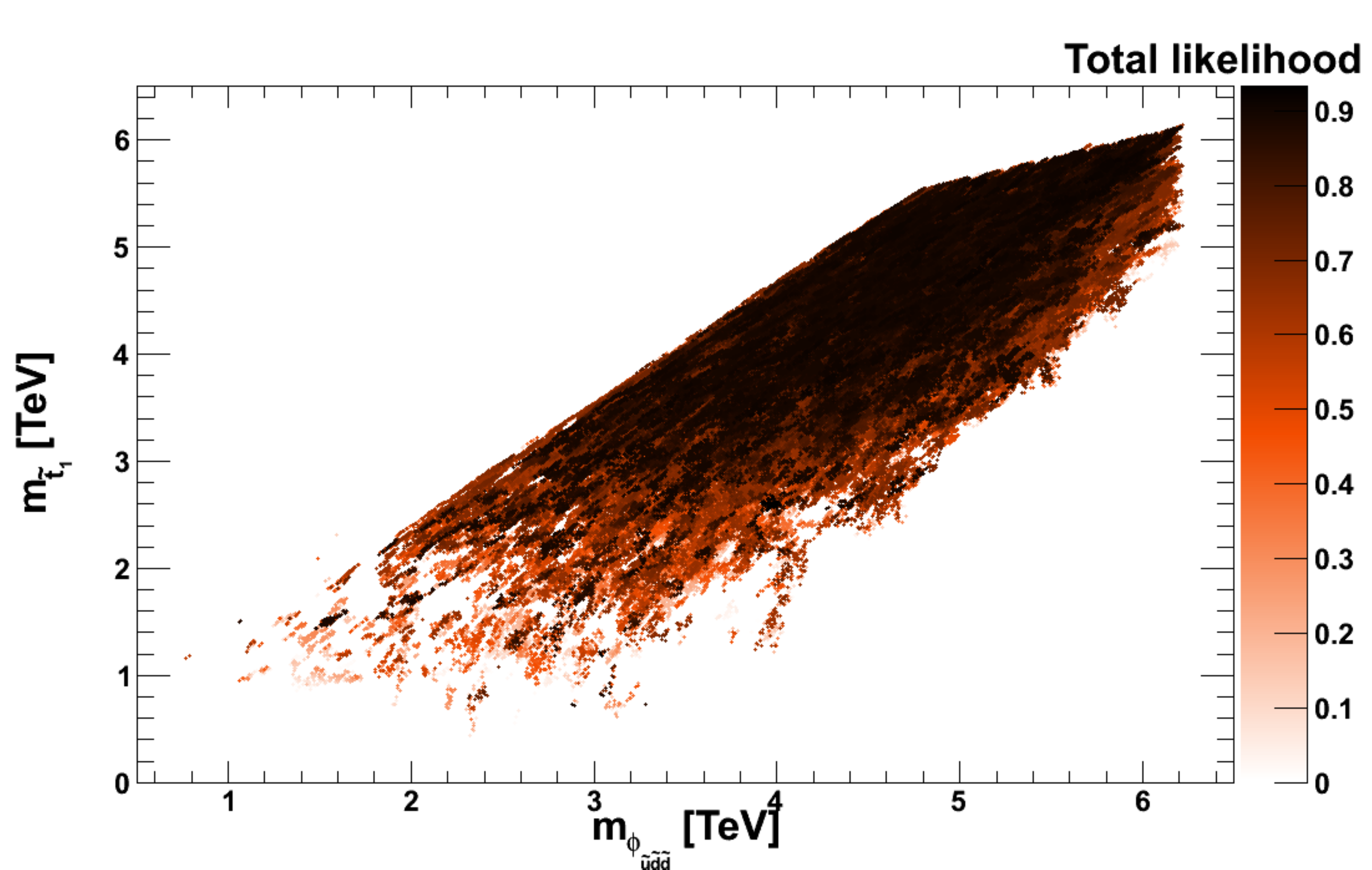}\\

\includegraphics[scale=0.23]{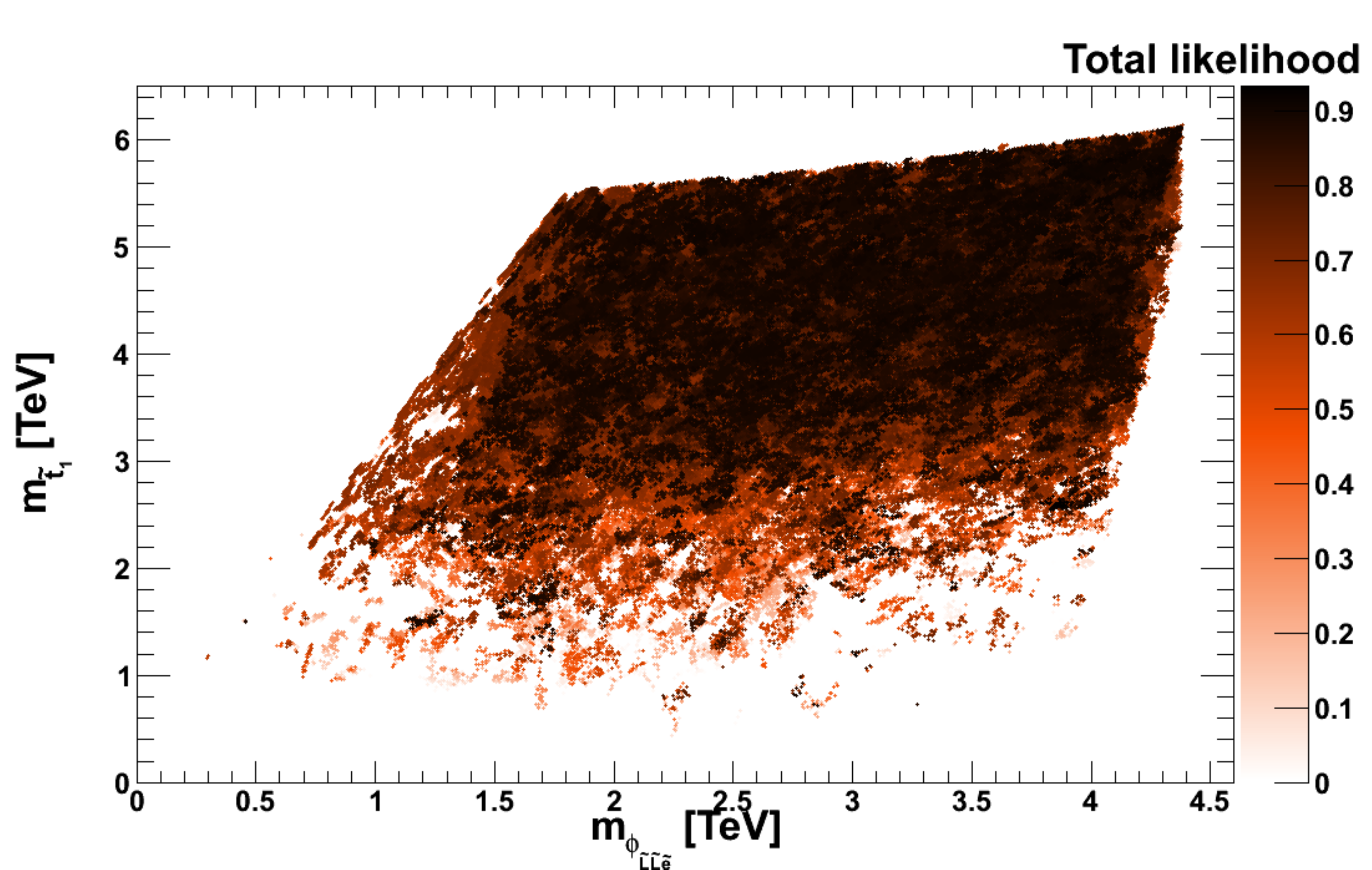}
\caption{The lightest stop mass $m_{\tilde{t}_1}$ versus the inflaton masses for $\widetilde u\widetilde d\widetilde d$ and $\widetilde L\widetilde L\widetilde e$, see Eq.~(\ref{masses}).}
\label{mstopmphi}
\end{figure}

Our previous scans of the NUHM2 parameter space have selected 
neutralinos with a high Higgsino fraction when the neutralino mass falls within the 0.6 and 1.2 TeV range. It is now interesting to check the prediction for the stop mass depending on the inflaton mass at TeV scale, see Fig.~(\ref{mstopmphi}). 
We find that in both inflation scenarios, the inflaton mass is above $500$ GeV and is associated with a very massive stop. For the $\widetilde{u}\widetilde{d}\widetilde{d}$ combination, the lightest stop mass is constrained to be within $m_{\phi} > m_{\tilde{t}_1} > m_{\phi}/3$. Scenarios with the lightest stops (namely $m_{\tilde{t}_1} \ls 2$ TeV) may offer a chance to probe the NUHM2 parameter space and thus a mean to determine the inflaton mass.


\begin{figure}[h]
\centering
\includegraphics[scale=0.23]{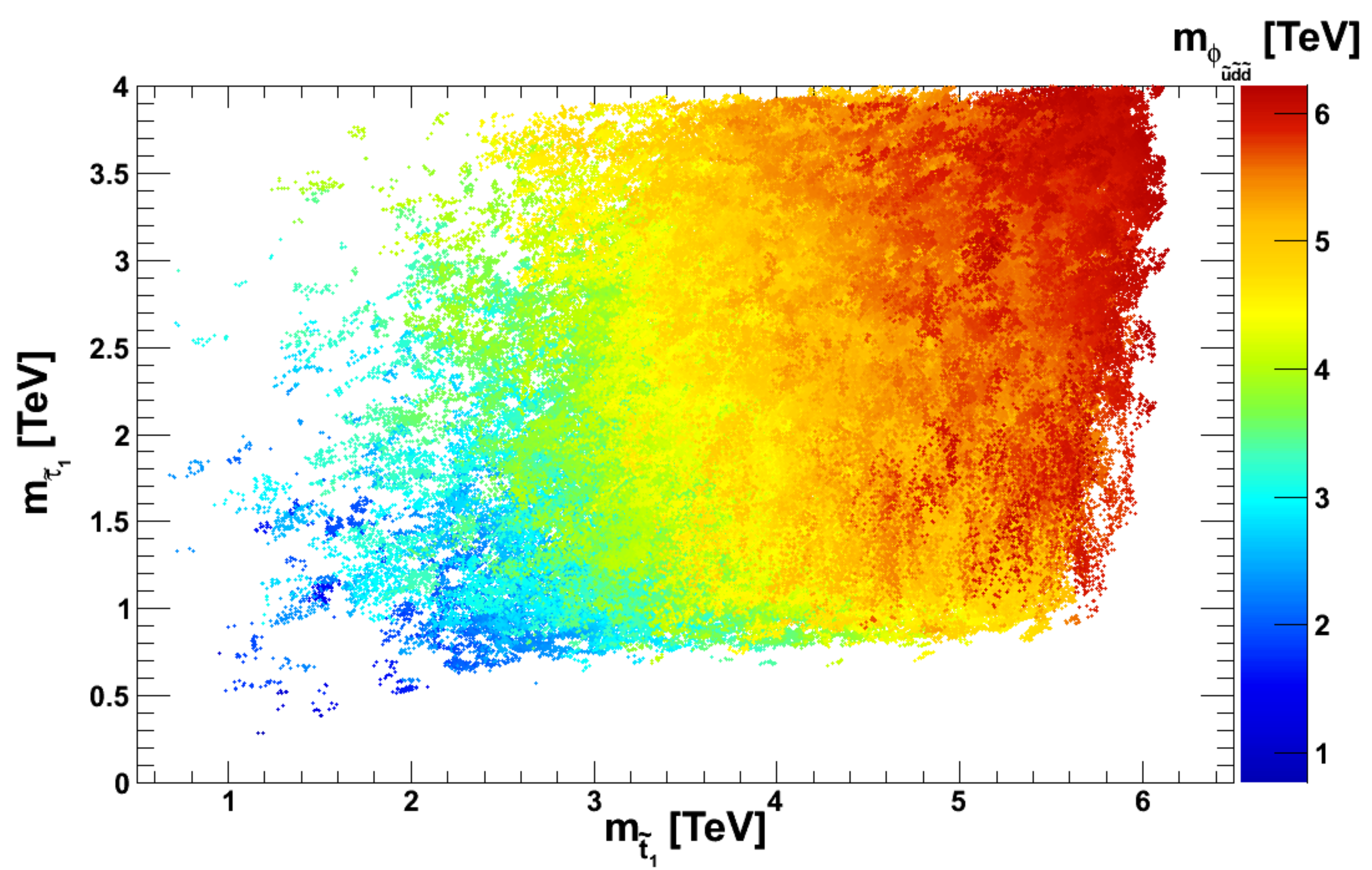}
\includegraphics[scale=0.23]{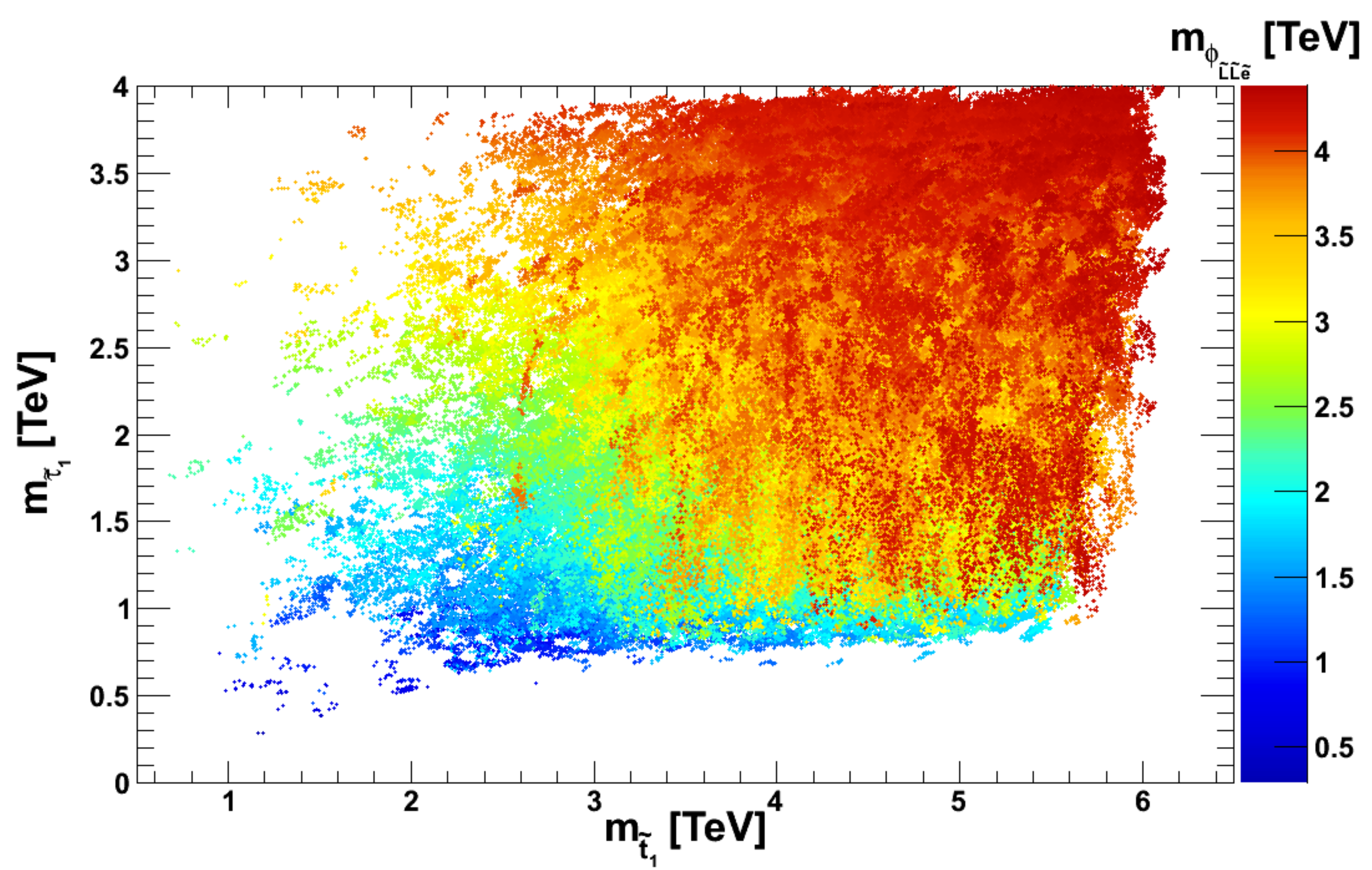}  
\caption{The correlation between stau mass, $m_{\tilde\tau_{1}}$, and the lightest stop mass, $m_{\tilde t_1}$. The color coding corresponds to the inflaton masses for $\widetilde u\widetilde d\widetilde d$ and $\widetilde L\widetilde L\widetilde e$.}
\label{staumasses}
\end{figure}

Such predictions have to be complemented by other observables, such as the stau mass Fig.~(\ref{staumasses}). The prediction differs depending on whether the inflaton correspond to the $\widetilde{u}\widetilde{d}\widetilde{d}$ or $\widetilde{L}\widetilde{L}\widetilde{e}$ inflation mechanism. For the $\widetilde{L}\widetilde{L}\widetilde{e}$  case, one finds that scenarios with 'light' inflaton (i.e. with a mass lower than 2 TeV) correspond to staus lighter than 2 TeV and stops lighter than 2-3 TeV. More generally there is a correspondence between the inflaton and the stau masses, whatever the value of the stop mass. This correlation between the stau and the $\widetilde{L}\widetilde{L}\widetilde{e}$ inflaton mass can be understood because the inflaton is of leptonic origin.  Similarly, for the $\widetilde{u}\widetilde{d}\widetilde{d}$ case, the inflaton mass is related to the stop mass but there is no constraint on the stau. Although such a 
feature can be easily understood given the nature of the inflaton, using LHC observables and searches for sparticles could provide a way to distinguish between the $\widetilde{u}\widetilde{d}\widetilde{d}$ and $\widetilde{L}\widetilde{L}\widetilde{e}$ scenarios. In addition, we find that staus in both scenarios can be lighter than $1$ TeV, thus offering another possible window for probing this model at LHC. Discovering a relatively light stau at LHC together with a specific stop mass would constrain the parameters of the model and thus provide a determination of the inflaton mass.

\begin{figure}[h]
\centering
\includegraphics[scale=0.23]{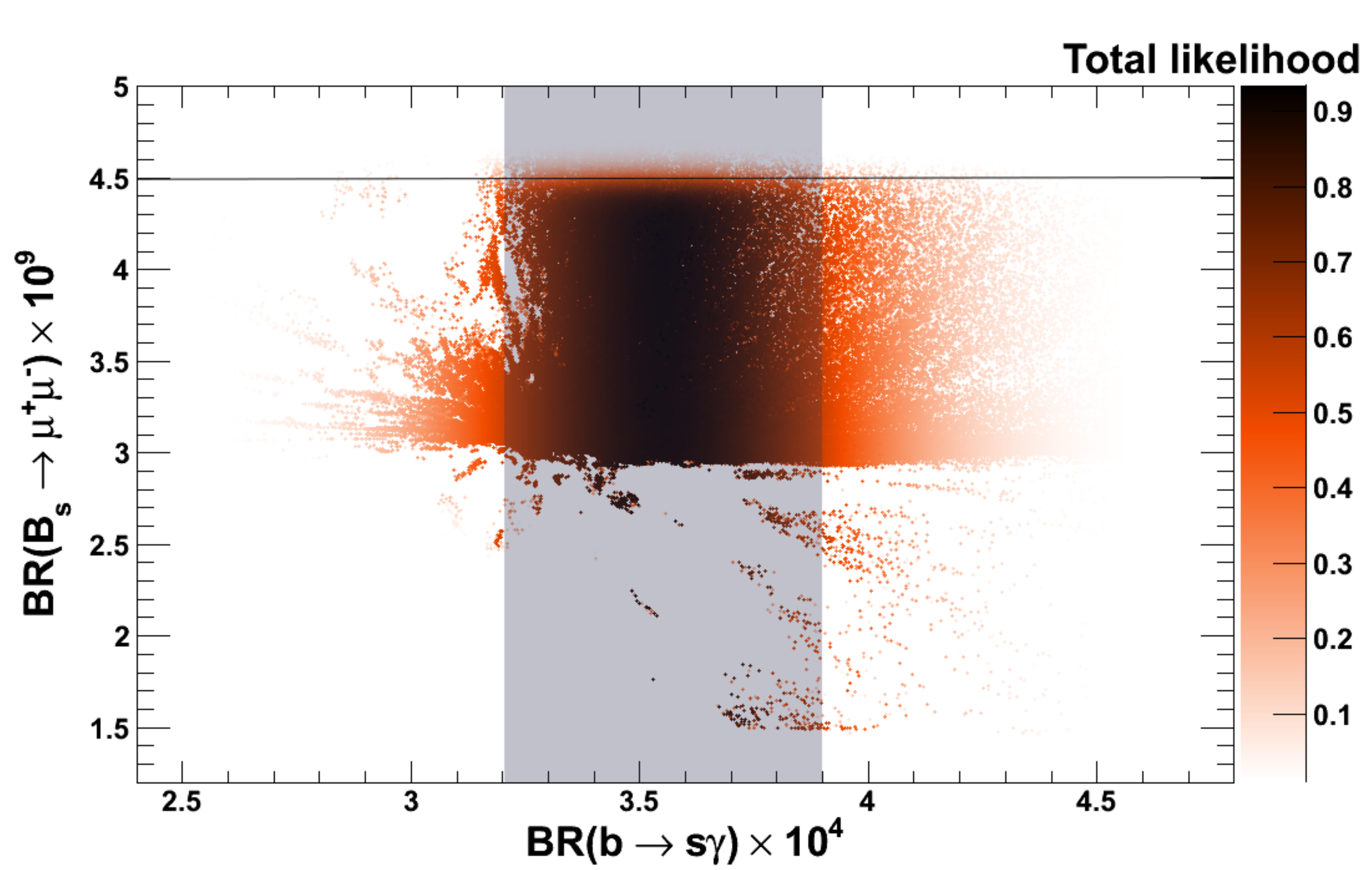}
\caption{The branching ratios of $B_s \rightarrow \mu^+ \mu^-$  and $b\rightarrow s \gamma$ are shown with the color coding corresponding to the likelihood. The shaded region shows points within $b\rightarrow s \gamma$ experimental and theoretical error bars. }
\label{bsmumu}
\end{figure}

Specific observables such as 
$B_s \rightarrow \mu^+ \mu^-$ and $b\rightarrow s \gamma$ are also interesting to 
consider. In particular, in Fig.~(\ref{bsmumu}), one can see that most of the scenarios which fall within the observed range of the $b \rightarrow s \gamma$ decay rate lead to a relatively large $B_s \rightarrow \mu^+ \mu^-$ branching ratio, basically within $3 \times 10^{-9}$ and $4.5 \times 10^{-9}$. Some scenarios are nevertheless excluded (i.e. with a contribution larger than $4.5 \ 10^{-9}$). This provides additional scope for detecting such scenarios at LHC since most scenarios are within the sensitivity of LHCb \cite{Aaij:2012ac}.


Finally, for completeness, we display the expected {\it spin-independent} elastic scattering cross section associated with these scenarios in a Xenon-based experiment. We juxtapose on this plot the limit obtained by the XENON100  experiment~\cite{Aprile2011} which is extremely robust regarding the relative scintillation efficiency $L_{eff}$ at this mass scale~\cite{Davis:2012vy}  (even though it may be affected by astrophysical uncertainties, 
see~\cite{McCabe:2011sr,Frandsen:2011gi} and uncertainties on quark coefficients of the nucleon), as well as the predicted limit for the XENON1T experiment.  

As one can see, most of the scenarios presented in this paper regarding NUHM2 are well below the present limit set by the XENON100 experiment and cannot be constrained for the moment. However the projected sensitivity for XENON1T indicates that it may be possible to probe NUHM2 parameters  in the forthcoming future if not already ruled out by the LHC.

 
\begin{figure}[h]
\centering
\includegraphics[scale=0.23]{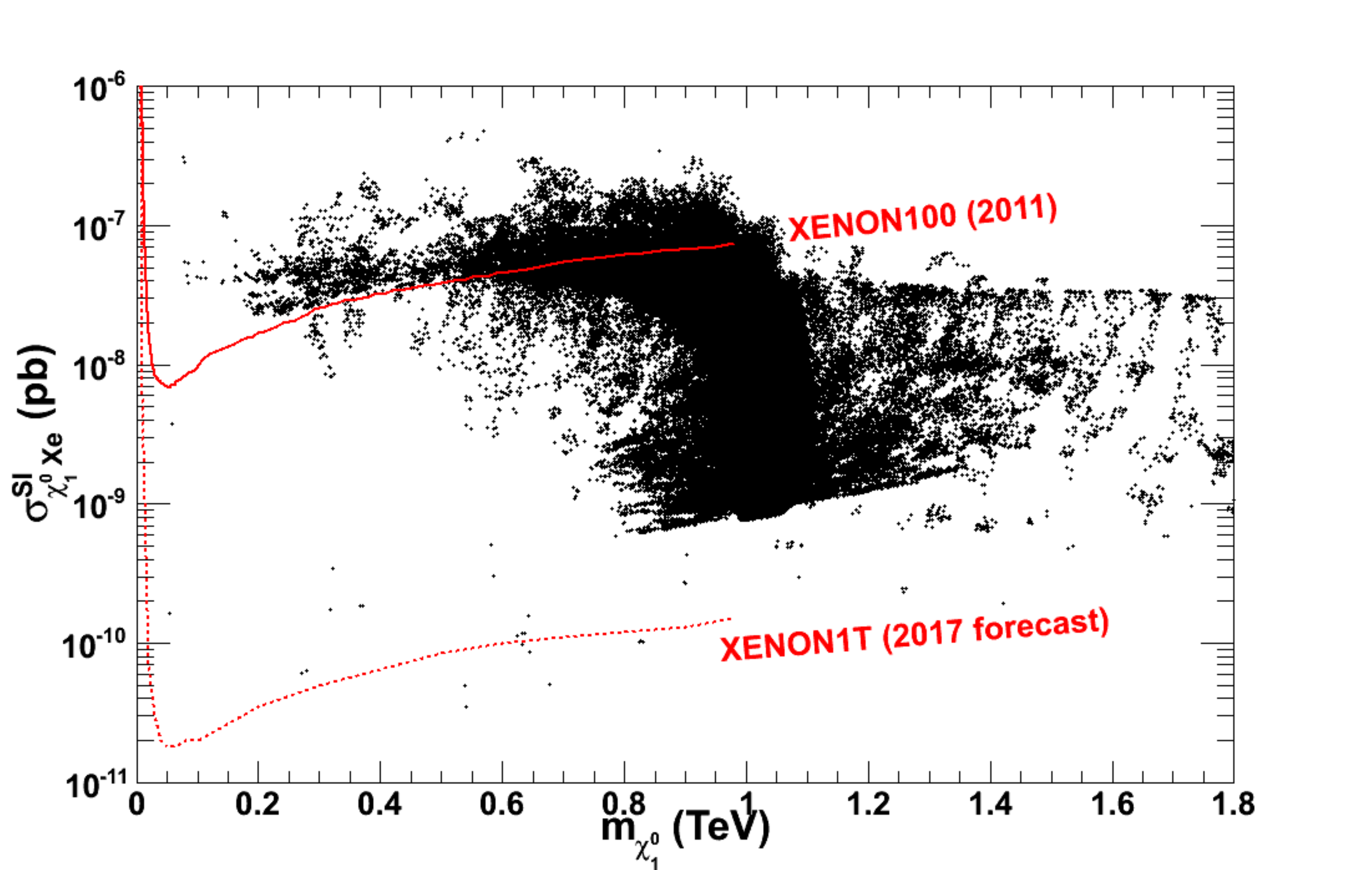}
\caption{The expected limit from XENON1T detector on the neutralino cross section ({\it spin-independent}) with respect to the neutralino mass.}
\end{figure}


\section{Conclusion and perspectives}

In this paper, we search for the regions of the NUHM2 (a variant of the MSSM with non-universal Higgs masses) parameter space which are compatible with the observed dark matter abundance (assuming that the neutralino is the dark matter candidate), the Higgs mass constraints from  LHC, and the constraints set on the inflationary potential to match the CMB constraints.

We have considered two inflaton candidates ($\widetilde{u}\widetilde{d}\widetilde{d}$ and $\widetilde{L}\widetilde{L}\widetilde{e}$) for which the 'high' scale of inflation $\phi_0$ is intimately tied up to the low scale physics at the LHC scale via the RGE, and which are compatible with the amplitude of the perturbations, $\delta_H=1.91\times 10^{-5}$ and the $2\sigma$  tilt in the power spectrum $0.934 \leq n_s\leq 0.988$~\cite{WMAP}. 

We used two methods. One consisted in finding benchmark points and the other one in performing a more complete scan of the parameter space by using a MCMC code. 
Our main conclusion is that for most configurations the $\widetilde{u}\widetilde{d}\widetilde{d}$  inflaton appears to be 'fairly light' but still heavier than $1$ TeV while the  $\widetilde{L}\widetilde{L}\widetilde{e}$ inflaton can be as light as 500 GeV.  In both cases however it is possible to find configurations in which both the staus and the stops are potentially within the reach of the LHC, thus indicating 
that sparticle searches at LHC could actually provide a mean to constrain the inflaton mass for some subset of the NUHM2 parameter space.  Such constraints would have to be cross correlated with the measurements of BR($B_s \rightarrow \mu^+ \mu^-$) and BR($b\rightarrow s\gamma$)   
since all the scenarios found in this paper have predicted values for these two branching ratios very close to the present experimental limits. Finally LHC constraints or potential hints could be enhanced by the results of the forthcoming dark matter direct detection experiments such as the XENON1T experiment.

As can be seen from Figs.~(\ref{fig3}) and (\ref{fig4}), hints of a TeV scale inflaton together with the precise measurement of the Higgs mass would  actually narrow down the scale of inflation.  
Combined with the Planck satellite measurements which is expected to constrain the range of the spectral tilt with a greater accuracy, one should actually be able to pin point both the scale of inflation $\phi_0$ and the corresponding mass $m_{\phi}$ at the scale of inflation, thus providing a window on extremely high energy physics which also complements the current observations from the CMB radiation.

To conclude, it is possible to embed inflation within MSSM. This interplay between inflation and dark matter provides an exciting prospect where inflationary paradigm can be tested by the Planck, LHC, along  with direct/indirect dark matter detection experiments.


\section{Acknowledgments}
We would like to thank Roubeh Allahverdi, Kaladi Babu, Arindam Chatterjee, and Qaisar Shafi for helpful discussions.
We thank Pran Nath for useful comments. The research of AM and EP are supported by the Lancaster-Manchester-Sheffield Consortium for Fundamental Physics under STFC grant ST/J000418/1. EP is supported by STFC ST/J501074. JD acknowledges the CMIRA 2011 EXPLO'RA DOC program of the French region Rh\^one-Alpes. AM acknowledges the award of the Royal Society grant that supported an India-UK seminar where this idea was discussed. CB thanks the same grant for their generous support. CB and AM also thank IUCAA (India) for their kind hospitality.


\end{document}